\documentclass[jgrga]{agu2001} % double column WITH FIGURES
\usepackage{graphicx} 
\usepackage{epstopdf}
\usepackage{setspace}
\usepackage{epstopdf}
\usepackage{amssymb}
\usepackage{url}
\usepackage{dcolumn}% Align table columns on decimal point
\newcommand{\be}{\begin{equation}}
\newcommand{\ee}{\end{equation}}
\newcommand{\ba}{\begin{eqnarray}}
\newcommand{\ea}{\end{eqnarray}}

\authorrunninghead{WERNER and SORNETTE}

\titlerunninghead{MAGNITUDE UNCERTAINTIES AND FORECASTS}

\authoraddr{M. J. Werner, Department of Earth and Space Sciences, and
        Institute of Geophysics and Planetary Physics, University of
        California, Los Angeles, California 90095, USA.
(werner@moho.ess.ucla.edu)}

\authoraddr{D. Sornette, ETH Z\"urich,
Chair of Entrepreneurial Risks,
Department of Management, Technology and Economics,
Kreuzplatz 5, CH-8032 Z\"urich, Switzerland and
Department of Earth and Space Sciences, and
Institute of Geophysics and Planetary Physics, University of
        California, Los Angeles, California 90095, USA 
(dsornette@ethz.ch)}

\begin{document}

\title{Magnitude Uncertainties Impact Seismic Rate Estimates, Forecasts and Predictability Experiments}

\today

\author{Maximilian J. Werner}
\affil{Department of Earth and Space Sciences, and Institute of
        Geophysics and Planetary Physics, University of California, Los
        Angeles, USA}

\author{Didier Sornette}
\affil{ETH Z\"urich, Chair of Entrepreneurial Risks,
Department of Management, Technology and Economics,
 Z\"urich, Switzerland, and
Department of Earth and Space Sciences, and
Institute of Geophysics and Planetary Physics, University of
        California, Los Angeles, USA}

\begin{abstract}

The Collaboratory for the Study of Earthquake Predictability (CSEP) aims to prospectively test time-dependent earthquake probability forecasts on their consistency with observations. To compete, time-dependent seismicity models are calibrated on earthquake catalog data. But catalogs contain much observational uncertainty. We study the impact of magnitude uncertainties on rate estimates in clustering models, on their forecasts and on their evaluation by CSEP's consistency tests. First, we quantify magnitude uncertainties. We find that magnitude uncertainty is more heavy-tailed than a Gaussian, such as a double-sided exponential distribution, with scale parameter $\nu_c=0.1 - 0.3$. Second, we study the impact of such noise on the forecasts of a simple clustering model which captures the main ingredients of popular short term models. We prove that the deviations of noisy forecasts from an exact forecast are power law distributed in the tail with exponent $\alpha=(a \nu_c)^{-1}$, where $a$ is the exponent of the productivity law of aftershocks. We further prove that the typical scale of the fluctuations remains sensitively dependent on the specific catalog. Third, we study how noisy forecasts are evaluated in CSEP consistency tests. Noisy forecasts are rejected more frequently than expected for a given confidence limit. The Poisson assumption of the consistency tests is inadequate for short-term forecast evaluations. To capture the idiosyncrasies of each model together with any propagating uncertainties, the forecasts need to specify the entire likelihood distribution of seismic rates.

\end{abstract}

\begin{article}

\section{Introduction}	

Earthquake prediction experiments such as the recently formed Collaboratory for the Study of Earthquake Predictability (CSEP) [{\it Jordan}, 2006] and the Working Group on Regional Earthquake Likelihood Models (RELM) [{\it Schorlemmer et al.}, 2007] aim to investigate scientific hypotheses about seismicity in a systematic, rigorous and truly prospective manner by evaluating the forecasts of models against observed earthquake parameters (time, location, magnitude, focal mechanism, etc) that are taken from earthquake catalogs. After the optimism of the 1970s followed by pessimism on earthquake prediction [e.g. {\it Geller},  1997], the lesson for the next generation of earthquake forecasters is clear: model formulation, calibration and hypothesis testing must be robust, especially with respect to data quality issues. 

Paleoseismology provides a good example of the importance of uncertainties in data analysis and hypothesis testing. {\it Rhoades et al.} [1994] showed that the hazard rate on the Pallett Creek segment of the San Andreas fault can vary by a factor of three depending on parameter estimates of any chosen model for the earthquake cycle, all consistent with the data. {\it Davis et al.} [1989] showed that parameter uncertainties and their continuous updating with time lead to vastly different probability forecasts of the Parkfield segment. In the context of renewal processes, {\it Sornette and Knopoff} [1997] showed that different distributions of inter-event times, which may all be compatible with the data, have drastically diverging implications for the conditional waiting time until the next earthquake. {\it Ogata} [1999, 2002] concluded that uncertainties in the occurrence times of historical quakes make differentiating between different renewal distributions for the hazard on one fault inconclusive. 

CSEP and RELM and future experiments have at their disposal earthquake catalogs of much higher quality than paleoseismic studies, of course. Nevertheless, these modern earthquake catalogs are ridden with their own observational uncertainties, biases, arbitrary conventions and spatio-temporally varying quality characteristics. While RELM acknowledges data problems and plans to simulate so-called ``modified observations" from the actual observations and their error estimates in order to test the forecasts against these alternative, potentially equally likely scenarios, their proposed statistics and tests were not shown to be sufficient to solve data quality issues at the various stages of the hypothesis testing process, in particular with respect to the generation of model forecasts. 

But models for short-term forecasts are typically quite sensitive to recent earthquake catalog data that the models are calibrated on. For instance, the seismic rate forecasts in popular clustering and aftershock models are exponentially dependent on main shock magnitudes. These models include the Epidemic Type Aftershock Sequence (ETAS) Model [{\it Ogata}, 1988; see {\it Helmstetter et al.}, 2006, for an implementation], Short Term Earthquake Probabilities (STEP) [{\it Gerstenberger et al.}, 2005], the model by {\it Reasenberg and Jones} [1989, 1994], the Epidemic Rate and State (ERS) model by {\it Console et al.} [2007] and the model by {\it Kagan and Knopoff} [1987]. Small errors in reported magnitudes can therefore have a large, detrimental effect on forecasts. Rather than entering at the evaluation phase, the magnitude uncertainties impact during the input or calibration stage - not accounted for by the current RELM/CSEP evaluation tests. To the best of our knowledge, no study has addressed the impact of data quality on seismic rate estimates, model forecasts and their evaluation. 

Here, we analyze how observational errors in magnitude estimates propagate to seismic rate estimates and forecasts in common aftershock models. While other data quality issues are certainly important, we focus here on magnitude uncertainties, which seem to be the most important source of noise for short term clustering models because of the models' exponential sensitivity on past magnitudes. However, because magnitude uncertainties are not routinely reported in earthquake catalogs, we first study the accuracy of magnitude estimates in section \ref{dm}. In section \ref{impact}, we then use a simple aftershock model containing the basic ingredients of most operational short term clustering models and investigate the impact of the magnitude errors on seismic rate estimates and forecasts. Finally, we conduct a numerical experiment in which we mimic the RELM/CSEP testing center to investigate the impact of noise on the evaluation of forecasts generated from noisy magnitude data. 

\section{Magnitude Uncertainties}
\label{dm}

\subsection{Different types of magnitudes and their errors}

Magnitude is a measurement unit originally introduced by Charles Richter and Beno Gutenberg during the 1930s at the California Institute of Technology to measure the ``size" of earthquakes. Different magnitudes measure different characteristics of seismic waves (or of earthquake-generated tsunamis or surface deformation), although all aim to quantify the ``size" of an earthquake. For example, the original Richter magnitude ($M_L$) uses the maximum amplitude of seismic waves and the time difference between P and S waves recorded on a Wood-Anderson seismograph along with correction factors for geometric spreading and attenuation. Body wave magnitudes ($m_b$) and surface wave magnitudes ($M_S$) are based on amplitudes of P and surface waves often near periods of $1$s and $20$s, respectively. Both may include corrections for focal depth, the period used, the attenuation and potentially a station correction for the inhomogeneity of the soil. 

The moment magnitude [{\it Kanamori}, 1977; {\it Hanks and Kanamori}, 1979] is based on the scalar seismic moment, which is the magnitude of the seismic moment tensor. The moment tensor, in turn, is a representation of the equivalent body forces that would produce the same radiated seismic wave pattern as the observed one. Under a point source approximation, the seismic moment is equal to the product of the rigidity of the material, the average displacement on the fault and the average fault area that slipped. The moment is a long period (zero frequency) measure of the total energy of the earthquake and in theory does not saturate. The moment magnitude has a clear physical interpretation, while the relationship of other magnitudes to earthquake source parameters is less (if at all) established. 

Many other definitions and conventions exist for magnitudes. For instance, the Advanced National Seismic System (ANSS) [{\it Benz et al.}, 2005] reports various magnitudes, including local (Richter-type) magnitudes, body wave magnitudes, moment magnitudes and coda duration magnitudes. Few earthquake catalogs homogeneously use the same magnitude type to measure earthquakes. The (Harvard) Centroid Moment Tensor (CMT) project catalog [e.g. {\it Ekstrom et al.}, 2005] is a rare exception of a relatively uniform global catalog. 

In principle, each magnitude type needs to be addressed separately to establish uncertainties. For non-physics-based magnitudes (i.e. all but the moment magnitude), this can be particularly difficult as they are conventions by definition and cannot be verified independently. However, even in these cases it is possible to establish estimates of the errors in the convention-based magnitude estimate. For instance, it is possible to analyze the effects of discretization of media and equations, of the measurement precision of seismometers, of the assumed velocity and attenuation models of the Earth (spherically symmetric, 3D, etc) and of the resolution of the inversion algorithm depending, e.g., on station coverage. We will use the term intra-magnitude uncertainties to refer to such individual magnitude error estimates. These uncertainties measure how close a magnitude estimate may be to its convention-based ``true" value. 

More fundamentally, the definition of an earthquake ``event" (and hence an associated magnitude) can be questioned. The identification of one earthquake seems inherently tied to the particular time, space and frequency resolution of the observer. {\it Kagan and Knopoff} [1981] constructed a branching model which mimics a continuous deformation flow. When a Green's function is applied to the deformation and a scale imposed, the resulting seismograms seem to show separate events. {\it Peng et al.} [2007] analyzed the properties of catalogs after large earthquakes and found by handpicking events from the waveforms that many more events are present than detected by catalog routines, providing further observational evidence to the idea that the scale of the observer may determine the definition of an event. In this article, we will assume that listed catalog magnitudes are nevertheless useful for extrapolating and forecasting seismic rates. This is the working assumption of all seismicity-based earthquake forecasting experiments. 

Earthquake prediction experiments such as RELM and CSEP use a so-called ``authorized data stream" for their ``natural laboratories" [{\it Schorlemmer et al.},  2007]. For California, this data stream is the ANSS catalog. Models accept the listed magnitudes to generate forecasts of future events, irrespective of the type of magnitude listed. The forecast validation is also performed against the listed magnitudes. Apart from the intra-magnitude uncertainties, one should therefore also consider the uncertainties of one particular magnitude estimate in relation to the magnitude that best forecasts future events (if there exists such a ``forecast magnitude"). For instance, the moment magnitude may be more relevant in predicting aftershocks than a body wave magnitude: A. Helmstetter (personal communication, 2007) observed unbroken scaling of the number of aftershocks with moment magnitude $\propto 10^{\alpha M_W}$ up to $M_W=9.3$ (the great December 26 2004 Sumatra-Andaman earthquake). Physical mechanisms for earthquake triggering might also be constrained by the knowledge of the ``forecast magnitude." 

But lacking this magnitude, we need to consider the uncertainties between the different types of magnitudes: the inter-magnitude uncertainties. We will study both  inter- and intra-magnitude uncertainties to get a sense of the scale of the uncertainties. We then use these error estimates to simulate noisy magnitudes and to study their impact on seismic rate estimates and forecasts in section \ref{impact}. 

\subsection{Intra-Magnitude Uncertainties}

Ideally, intra-magnitude uncertainties are reported by earthquake catalogs based on knowledge about the seismic instruments and the inversion algorithm. Unfortunately, such information is often lacking in catalogs. A rare exception is provided by the Northern California Seismic Network (NCSN) catalog 
operated by the U.S. Geological Survey (USGS) and the Berkeley Seismological Laboratory at UC Berkeley. We study the reported uncertainties in section \ref{NCSNnoise}. 

A simple alternative to studying these errors independently is to compare the magnitude estimates for the same event from different networks and inversion algorithms, e.g. from different catalogs. While one cannot assume that one measure is correct and indicative of the error in the other, one can make some inferences, especially if the catalogs seem uniform and trustworthy (established, e.g., by verifying completeness, stability and other known statistical properties). We therefore study the differences in moment magnitudes as reported by two relatively well-studied and trusted catalogs: the (Harvard) Centroid Moment Tensor (CMT) catalog [{\it Dziewonski et al.}, 1981; {\it Dziewonski and Woodhouse}, 1983; {\it Ekstrom et al.}, 2005] and the USGS moment tensor (MT) catalog [{\it Sipkin}, 1986, 1994].

\subsubsection{Moment Magnitude Uncertainty From the (Harvard) CMT and USGS MT Catalogs}
\label{MwNoise}

{\it Sipkin} (1986, Figure 7) compared the CMT scalar moment tensor with the USGS-MT scalar moment tensor. He generally found comparable values, but CMT moments were smaller by a factor of two for small events and slightly larger for the largest events. The scatter shows deviations of 0.25 in logarithmic moment units. {\it Helffrich} (1997) compared the moment tensor solutions provided by three organizations: the Harvard group, the USGS and the Earthquake Research Institute of the University of Tokyo. He found a standard deviation of $0.21$ in $\log_{10}$ moment units between the three data sets. He also showed that the moment estimates systematically improved (converged) for deeper events. {\it Kagan} (2003) found that the differences in moment magnitude estimates of matched events reported by the Harvard CMT and the USGS MT catalog have a standard deviation of $0.08$ and $0.12$ for deep and shallow events, respectively. If both catalogs contain an equal amount of noise, then the standard deviation of each estimate is equal to 0.05 and 0.08. {\it Kagan} (2002) concluded that the standard deviations for moment magnitude estimates in California was 0.08, while conventional (first motion) catalogs provided less accurate estimates with a standard deviation equal to 0.23. Since we are interested in simulating noisy magnitudes to study their impact on seismic rate forecasts and prediction experiments, we update and expand these analyses of moment magnitude to determine the entire distribution. 

We used the Harvard CMT catalog from 1 January 1977 until 31 July 2006, which contains 25066 events above $M_W \geq 3$ and wrote an algorithm to match its events with the USGS MT catalog from January 1980 until 31 July 2006, which contains 4952 events above $M_W \geq 3$. Both catalogs are available from \url{http://neic.usgs.gov/neis/sopar/}. Neither catalog contains events between 1 December 2005 and 31 March 2006. But since we are not interested in temporal properties of the catalogs, this gap should not bias our results. 

We consider two listings from the two catalogs to refer to the same event if they are separated in time by less than 1 minute and in space by less than 150 km. {\it Kagan} (2003) used the same definitions. In agreement with his findings, the matches are quite robust with respect to space but less robust with respect to the condition on time. 

Using these conditions, we match $4923$ pairs of events. Only 29 events listed in the USGS MT catalog cannot be matched with events in the Harvard CMT catalog. Of these, 5 events pass the time requirement but fail the spatial condition. Thus one might suspect extreme errors in locating the events. However, increasing the spatial limit up to 1000 km does not change the results, i.e. the differences do not seem to be due simply to large location errors. They seem to be events listed in the USGS MT catalog that are entirely absent from the Harvard CMT catalog. Most of the other 24 events listed in the USGS MT that could not be matched with events Harvard CMT seem to be events in complex aftershock sequences where the identification of single events may sensitively depend on certain different network characteristics and on choices in the two computer algorithms. Vice versa, many events listed in the Harvard CMT are absent from the USGS MT, often due to a lower detection threshold in the Harvard CMT catalog. 

\begin{figure}[!ht]
\begin{center}
\includegraphics[width=16cm]{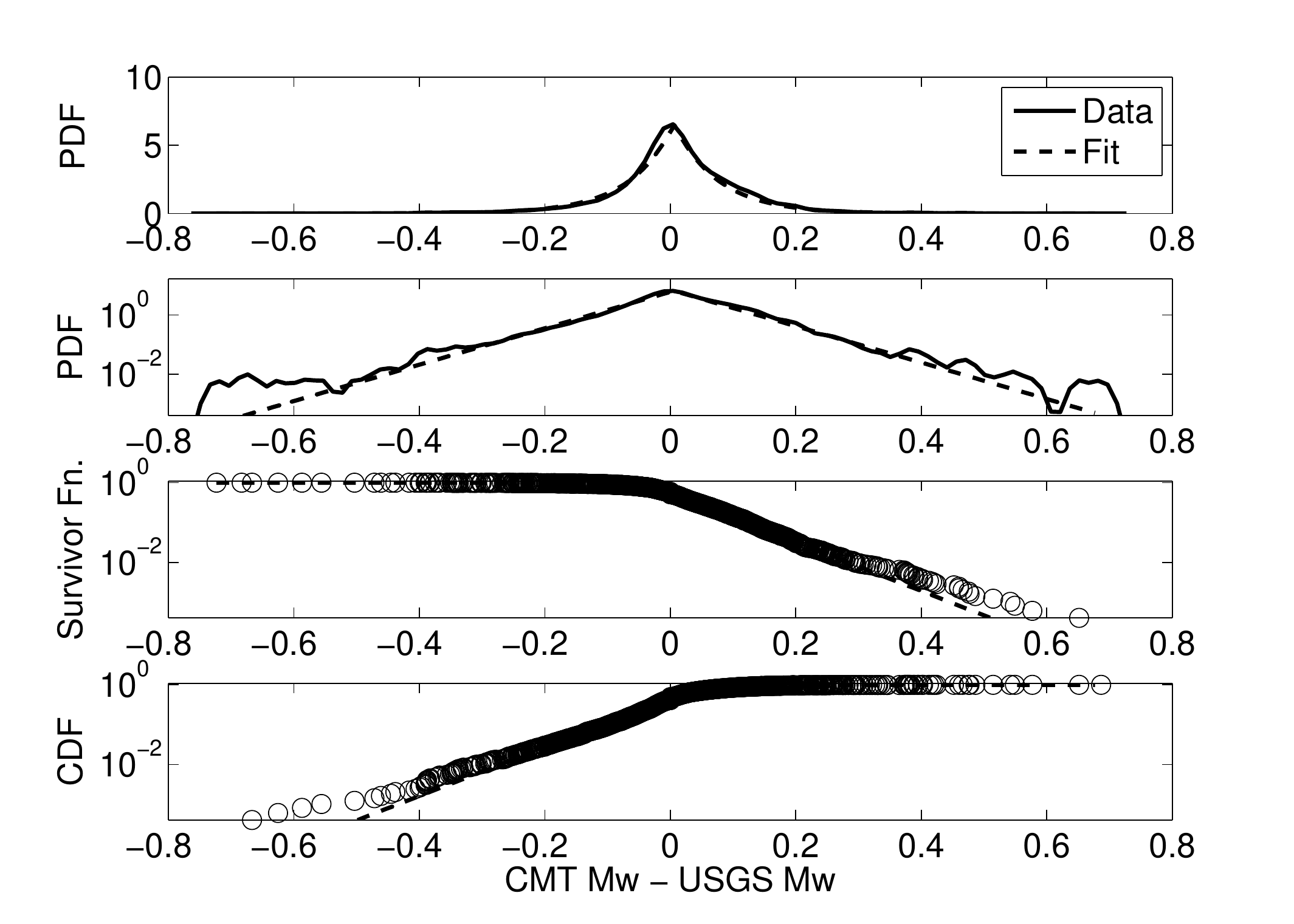}
\end{center}
\caption{\label{Mw} Estimating intra-magnitude uncertainty by comparing moment magnitude estimates for the same event from the Harvard CMT and the USGS MT catalogs. {\bf a)} Fixed kernel density estimate (solid) of the probability density function of the differences in moment magnitudes and maximum likelihood fit (dashed) of a Laplace (double-sided exponential) distribution given by equation (\ref{noise}) with scale parameter $\nu_c=0.07$. {\bf b)} Same as a) but in semi-logarithmic scale. {\bf c) } Semi-logarithmic plot of the survivor function (complementary cumulative distribution function). {\bf d) } Semi-logarithmic plot of the cumulative distribution function. 
}
\end{figure}

For matched events, we calculate the moment magnitude $M_W$ from the scalar moment $M_0$ (in Newton-meter) using the relation $M_W= 2/3 \ \log_{10} (M_0)-6$ [{\it Kagan}, 2003] and analyze the differences in $M_W$ between Harvard CMT and USGS MT estimates. Figure \ref{Mw} shows the distribution of the differences in the moment magnitude estimates from the two different catalogs. Figure \ref{Mw}a) shows a fixed kernel density estimate [e.g. {\it Izenman}, 1991] of the probability density function (pdf) of the magnitude uncertainties (solid). Figure \ref{Mw}b) shows the same data in a semi-logarithmic plot. Figures \ref{Mw}c) and d) show semi-logarithmic plots of the survivor function and cumulative distribution function, respectively, in order to emphasize both tails.

We performed a maximum likelihood fit of the data to a Laplace distribution (a double-sided exponential distribution), defined by: 
\be
p_{\epsilon} (\epsilon ) = { 1 \over 2 \nu_c} e^{\left(  - { |\epsilon - \langle \epsilon \rangle | \over \nu_c}   \right)}
\label{noise}
\ee
where $\nu_c$ is the scale (e-folding) parameter indicating the strength of the noise and $\langle \epsilon \rangle$ is a shift parameter equal to the median and mean of the distribution. The maximum likelihood estimate of $\langle \epsilon \rangle=0.006$ is given by the median and essentially indistinguishable from zero. To estimate the scale parameter $\nu_c$ (e-folding scale), we then took absolute values of the deviations from the median and fit the resulting positive data set to an exponential using maximum likelihood. We find that $\nu_c=0.07$ for the entire data set. The dashed lines in all plots of Figure \ref{Mw} correspond to this maximum likelihood fit (using $\langle \epsilon \rangle=0.006$ and $\nu_c=0.07$). The fit approximates the data well in the body, but underestimates the tails as can be seen in Figure \ref{Mw}b), c) and d). 

\begin{figure}[!ht]
\begin{center}
\includegraphics[width=16cm]{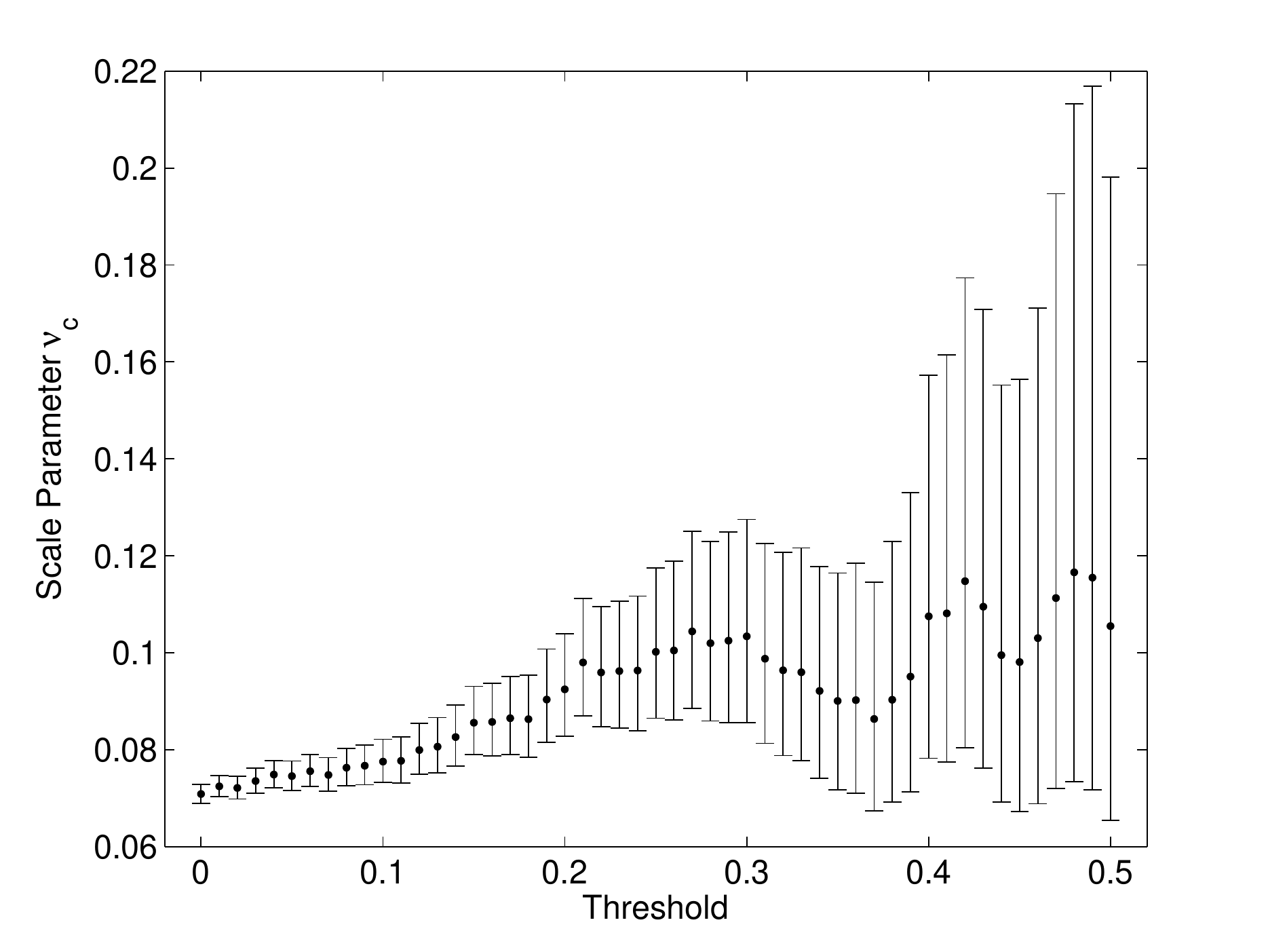}
\end{center}
\caption{\label{nu} Estimates of the e-folding scale parameter $\nu_c$ of the Laplace (double-sided exponential) distribution of equation (\ref{noise}) as a function of the threshold above which the data is fit. $\nu_c$ increases from $0.07$ to about $0.1$ due to fatter-than-exponential tails before starting to fluctuate more strongly due to finite sample effects. Error bars show 95$\%$ confidence intervals. 
}
\end{figure}

To estimate the effect of the tails which are fatter than exponential, we determined the scale parameter $\nu_c$ as a function of the threshold above which we fit the distribution. We determined the median of the data (corresponding to the threshold ``0"), took absolute values of the data and performed a maximum likelihood fit to an exponential distribution. We then increased the threshold (with respect to the median) and re-calculated the scale parameter for each threshold value. Figure \ref{nu} shows the resulting maximum likelihood estimates of $\nu_c$ as a function of the threshold. The error bars show estimated 95 percent confidence bounds. Confirming the presence of the fat tails, the estimated e-folding scale $\nu_c$ increases with the threshold from about $0.07$ to about $0.1$. 

The distribution of the differences between magnitude estimates is not the same as the distribution of the individual magnitude uncertainties in one estimate (see the next section \ref{NCSNnoise} for such a direct estimate). To obtain individual uncertainties, one can assume, for instance, that both uncertainties are identically and independently distributed (i.i.d.). In this case, the distribution of the differences can be assumed to be the convolution of the two individual distributions. In the case of Gaussian distributions, the convolution is also a Gaussian with variance equal to the sum of the individual variances. Unfortunately, the Laplace distribution is not endowed with the same property. The difference between two Laplace distributed random variables is not exactly another Laplace distributed random variable.

While we cannot determine the distributions of the individual uncertainties exactly, we can, however, constrain them. For instance, they cannot be Gaussian for the above-mentioned property that their differences would be Gaussian. The presence of exponentially or even more slowly decaying tails indicates that the individual uncertainties have at least exponentially decaying tails. For instance, to obtain a distribution for the difference of two random variables which looks approximately Laplace with scale parameter $0.1$, we found that using Laplace distributions for the individual variables requires them to have an e-folding parameter equal to $0.07$. We even found that the convolution of power law distributions was able to produce the observed tails more accurately. In summary, the individual uncertainty distributions must have tails that decay as slowly as or more slowly than exponential. 

\subsubsection{Intra-Magnitude Uncertainties Reported by the NCSN}
\label{NCSNnoise}

In the context of the CSEP earthquake prediction experiment, the important intra-magnitude uncertainties should be evaluated in California and in the regions where natural laboratories are being established around the world (e.g. Europe, New Zealand, West Pacific). For California, the earthquake catalog that provides the data for the RELM and CSEP experiments is the ANSS composite catalog. 

Many regional networks feed their data into the ANSS composite catalog. It is essentially a computer program with rules for merging the data [e.g. {\it Oppenheimer}, 2007]. The ANSS assigns so-called ``authoritative" regions to each seismic network, meaning that in those regions only data from its designated network is accepted into the composite catalog. The Northern California Seismic Network (NCSN) fulfills this role for northern California. The earthquake data that is passed on to the ANSS by the NCSN in turn comes from two sources: the Berkeley Seismological Laboratory at the University of California, Berkeley (UCB) and the USGS at Menlo Park. 

The UCB reports mainly moment magnitudes and local magnitudes, while the USGS reports coda duration magnitudes. Their policy for merging into ANSS is: moment magnitude supercedes local magnitude supercedes coda duration magnitude (David Oppenheimer, 2007, personal communication). UCB does not report uncertainties, but the moment magnitude is believed
to be the most stable with uncertainties around $0.1$, while the scatter
in local magnitudes is strongly affected by radiation pattern (approximately 0.15 magnitude units) and can be as large as $\pm 0.5$ (Margaret Hellweg, 2007, personal communication). 

The USGS, on the other hand, does provide uncertainties to the ANSS catalog, based on inversions by the Hypoinverse program, written by Fred W. Klein of the USGS [{\it Klein}, 2002]. The Hypoinverse code ``processes files of seismic station data for an earthquake (like P-wave arrival times and seismogram amplitudes and durations) into earthquake locations and magnitudes." The summary magnitude for an event is the weighted median of the station magnitudes. Each station can report a magnitude for an event if its user-specified weight is non-zero. The final reported magnitude is the value for which half of the total weights are higher and half lower. 

The measure of uncertainty reported by the Hypoinverse program, available from the NCSN in its hypoinverse format output and from the ANSS in its ``raw" format, is the Median Absolute Difference (MAD) between the summary magnitude and the magnitudes from the other reporting stations. 

The MAD value measures the variability of the magnitude estimates across several stations. It therefore probes the assumed velocity, attenuation and geometrical spreading models and the differences in station properties (e.g. frequency response, gain etc). Systematic biases due to other reasons in the magnitude inversion, however, cannot be captured by this measure. For instance, phase picking 
or instrument calibration may be systematically biased. Furthermore, only one measure (the median) of the entire distribution of the magnitudes from the different stations calculated for one event is reported. If the distributions are heavy-tailed, then the median may give a false sense of good measurement
in the presence of large variability. 

Nevertheless, some inference can be made about the accuracy of determined magnitudes. We collected all earthquakes in the NCSN's authoritative region of the ANSS, defined by a polygon with the following latitude and longitude coordinates: $\{$ 34.5, -121.25, 37.2167, -118.0167, 37.75, -118.25,  37.75, -119.5, 39.5, -120.75, 42.0, -121.4167, 42.0, -122.7, 43.0, -125.0, 40.0, -125.5, 34.5, -121.25 $\}$. We selected data from 1/1/1984 to 31/12/2006 (inclusive) above a magnitude threshold $m_{th}=3$. The data with MAD values can be retrieved from the website  \url{http://www.ncedc.org/ncedc/catalog-search.html} by selecting output in the ``raw" format. This yielded 6125 earthquakes. But, as already mentioned, only the USGS reports MAD values for its magnitudes, leaving 3073 events.

\begin{figure}[!ht]
\begin{center}
\includegraphics[width=16cm]{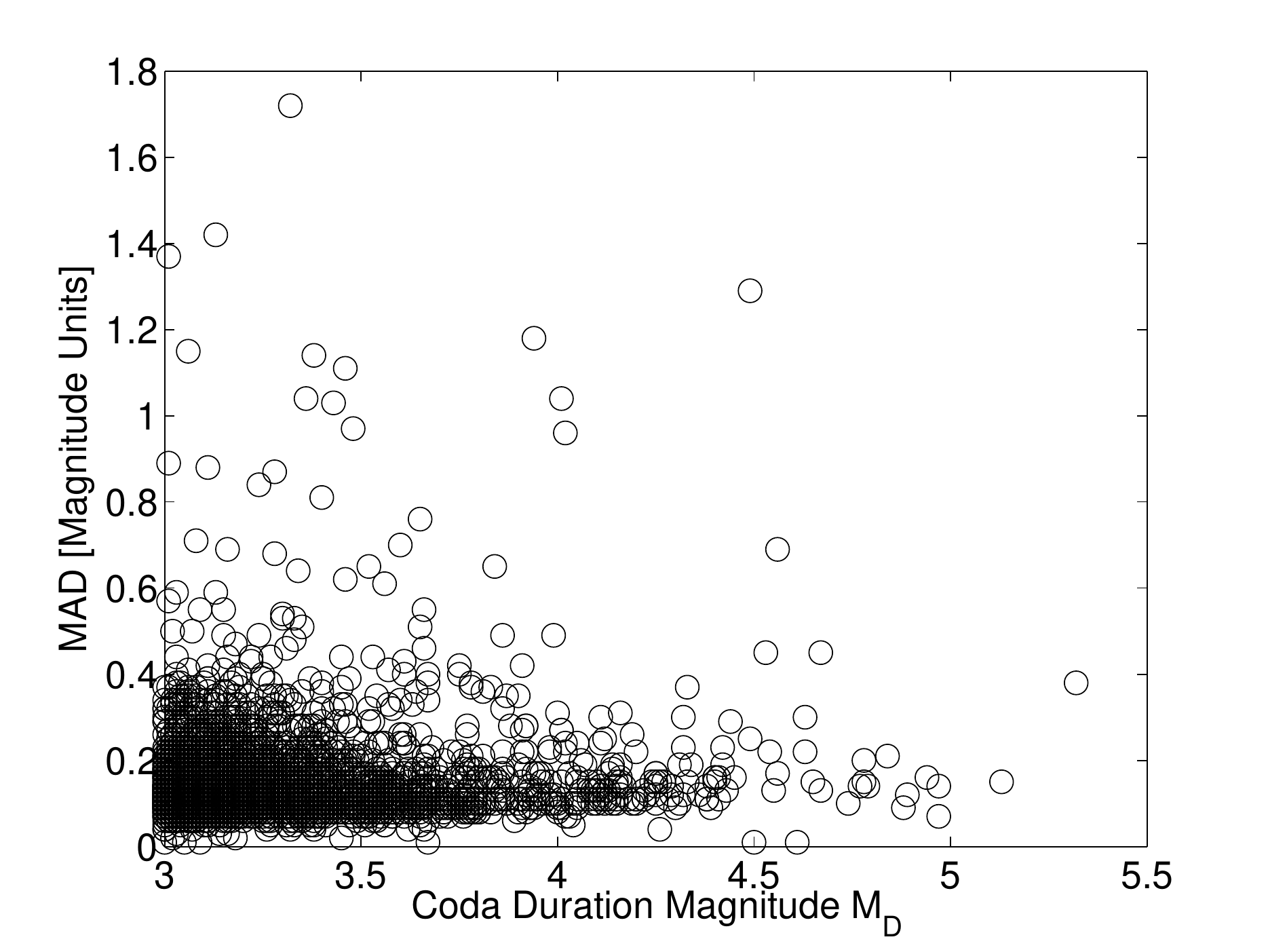}
\end{center}
\caption{\label{MAD1} Median Absolute Differences (MAD) versus their associated coda duration magnitudes as reported by the NCSN in its authoritative region of the ANSS composite catalog. MAD values measure the variability of the magnitude estimates for the same event from different stations as computed from the HYPOINVERSE program of the USGS. 
}
\end{figure}

In Figure \ref{MAD1}, we show a scatter plot of the MAD versus their associated duration magnitudes. To test for a decrease of MAD with increasing magnitude, we divided the magnitudes into bins of size $0.5$ and calculated the mean MAD for each bin. In the range $3.5<m<4.0$ (2420 events), the mean MAD was 0.15; for $4.0<m<4.5$ (372 events), the mean was 0.16; for $4.5<m<5.0$ (22 events), the mean was 0.20. The bin $5.0<m<5.5$ had mean 0.27 but counted only 2 events. Rather than a decrease of MAD with increasing magnitude, we see some evidence for an increase.

\begin{figure}[!ht]
\begin{center}
\end{center}
\includegraphics[width=16cm]{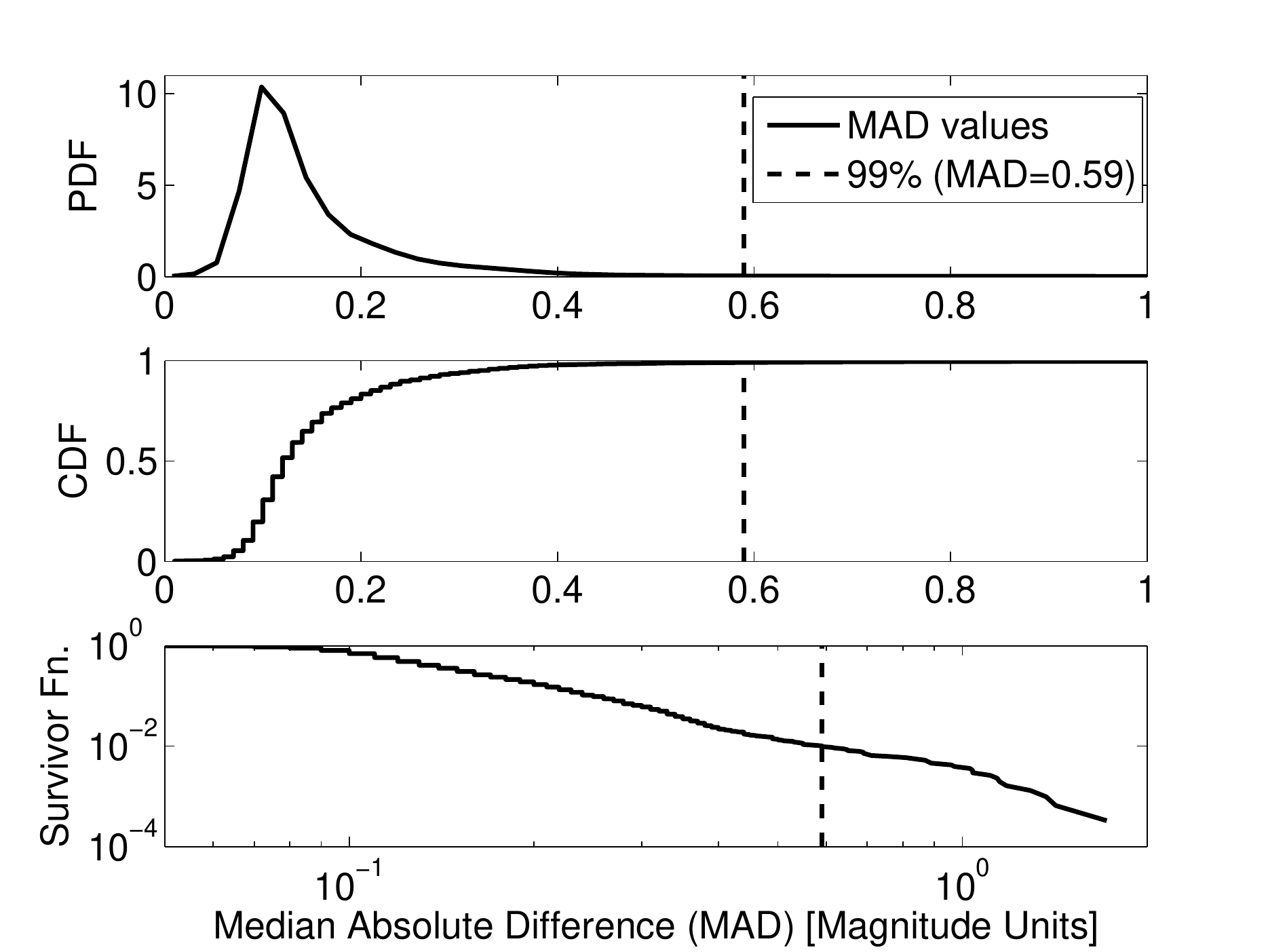}
\caption{\label{MAD2} Median Absolute Differences (MAD) reported by the NCSN in its authoritative region of the ANSS composite catalog. Top: kernel density estimate of the probability density function (pdf). Middle: cumulative distribution function (CDF). Bottom: survivor function plotted on logarithmic axes. The dashed line at MAD=0.59 corresponds to the $99$th percentile of the distribution. 
}
\end{figure}

Figure \ref{MAD2} shows the kernel density estimate of the probability density function (pdf) of the MAD values, the cumulative distribution function (cdf) and the survivor function. For reference, we also included the $99\%$ confidence limit (the MAD value for which $99\%$ of all reported MAD values are smaller). While the mean of the values was $0.15$ and the standard deviation $0.1$, the $99\%$ confidence limit was only reached at $0.59$. That the distribution is strongly non-Gaussian can also be seen from the bottom panel. The tails decay more slowly than an exponential, indicating that outliers occur relatively frequently. Indeed, the maximum MAD value was $1.72$. 

\begin{figure}[!ht]
\begin{center}
\includegraphics[width=16cm]{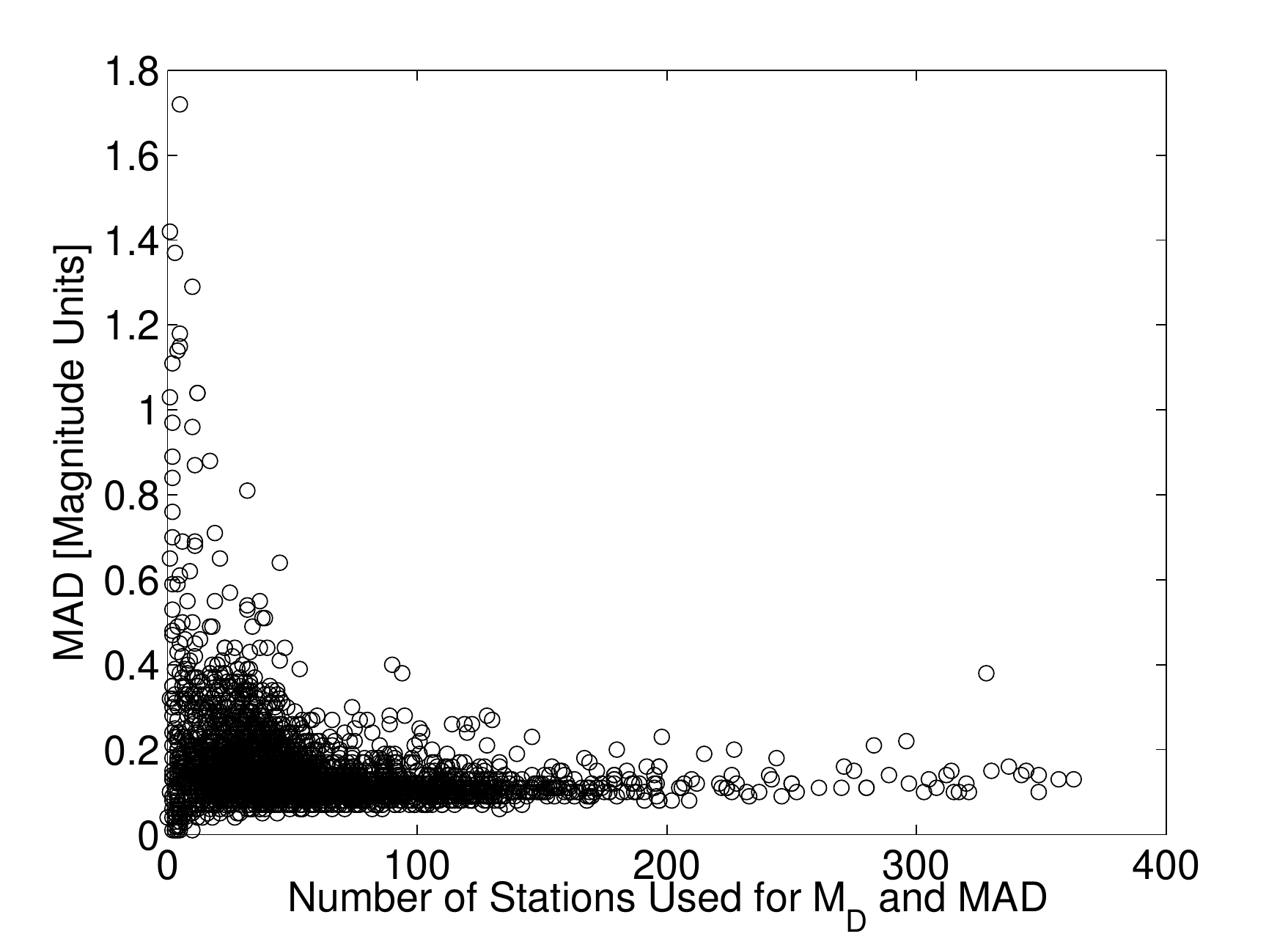}
\end{center}
\caption{\label{MAD3} Median Absolute Differences (MAD) reported by the NCSN in its authoritative region of the ANSS composite catalog as a function of the number of stations involved in the computation of the magnitude and the MAD value.  
}
\end{figure}

Figure \ref{MAD3} shows a scatter plot of the MAD values versus the number of stations involved in the computation of each coda duration magnitude and its MAD value. When the number of stations involved is very small, we see large scatter - very small MAD values of less than 0.1 and very large values above 0.5. On the other hand, as more stations are involved, the smallest MAD values increase to about 0.1. This may indicate that MAD values less than $0.1$ are due to too few stations involved in the computation and probably unreliable. (At the same time, we note that a $M_D=5.32$ event with MAD $0.38$ was recorded by 328 stations, suggesting that large MAD values are real.)

When the number of stations registering an event is small, this presumably implies that the event is small and/or remote. Given that many events are located by less than 10 stations, we may have detected evidence that the ANSS is not complete in the authoritative region of the NCSN down to $m=3$, because even some $M_D \sim 5$ events are constrained only by few stations.

Finally, it is difficult to interpret the group of large MAD values reported when few stations are involved. Perhaps the events belong to a particular group defined by a region or period with special properties that are not well modeled by the Hypoinverse program. 

\subsection{Inter-Magnitude Uncertainties}

As was mentioned before in this article, models in earthquake prediction experiments indiscriminately use whatever type of magnitude is listed in the authorized data stream. Given the random and systematic differences between the magnitudes, the relevant magnitude uncertainties in the context of forecasting are in fact inter-magnitude uncertainties. Many studies have investigated the relation of one magnitude scale to another and their random scatter. We review some of those here. We then analyze two data sets: (i) the differences between the CMT moment magnitudes and their corresponding body or surface wave magnitudes from the catalog provided by the Preliminary Determination of Epicenters ({\it PDE}, 2001) (section \ref{inter1}); and (ii) the differences between duration and local magnitudes in the NCSN (section \ref{inter2}). 

{\it Sipkin} (1986, Figures 1 and 2) compared the body wave magnitudes $m_b$ and surface wave magnitudes $M_S$ from the PDE catalog with early USGS scalar moments. Body wave magnitudes were scattered by up to one magnitude unit for the same value of the seismic moment, while surface wave magnitudes were less scattered but also substantial. {\it Dziewonski and Woodhouse} (1983, Figures 14a and 14b) compared the CMT moment with the PDE body wave magnitudes and found a similar amount of scatter.

{\it Harte and Vere-Jones} (1999) compared the properties of the local New Zealand catalog with the data from the PDE catalog, made available by the National Earthquake Information Center (NEIC). They concluded that the differences between the PDE's $m_b$ and the local catalog's $M_L$ were random for shallow events, but up to 1 unit of magnitude in size. For deeper events, $m_b$ was systematically smaller than the local $M_L$. 

{\it Kuge} (1992) found a systematic bias in the body wave magnitude $m_b$ reported by the International Seismic Center (ISC) and the converted seismic moment taken from the Harvard CMT catalog for intermediate and deep earthquakes in Japan. He computed a ``theoretical" $m_b$ based on the CMT seismic moment and regression relationships between the moment and NEIS (National Earthquake Information Service) and ISC body wave magnitudes [{\it Giardini}, 1988]. The systematic differences were on the order of 0.2 to 0.3 units. 

{\it Patton} (2001) investigated the transportability of the Nuttli magnitude scale based on 1-Hz Lg waves to different regions of the world. He routinely found differences of different types of body wave magnitudes on the order of 0.3 magnitude units. 

{\it Kagan} (2003) found that $m_b$ to $M_W$ conversions could result in scatter with a standard deviation of $0.41$. He also concluded that converting conventional magnitudes into moment magnitude leads to uncertainties which are three to four times the errors in moment magnitude estimates ($0.05$ to $0.08$).

\subsubsection{Moment Magnitude vs Body and Surface Wave Magnitude in the CMT and PDE Catalogs}
\label{inter1}

The Harvard CMT catalog calculates moment magnitudes when it receives notification of a large event either from the NEIC via the PDE system or from the ISC. We compared the seismic moments of the Harvard CMT with the original PDE body ($m_b$) and/or surface wave ($M_S$) magnitudes. That large systematic differences exist between these magnitudes is well-known. Here, we look at the differences between the Harvard $M_W$ and the PDE $m_b$ and $M_S$ estimates to evaluate their scale. 

\begin{figure}[!ht]
\begin{center}
\includegraphics[width=16cm]{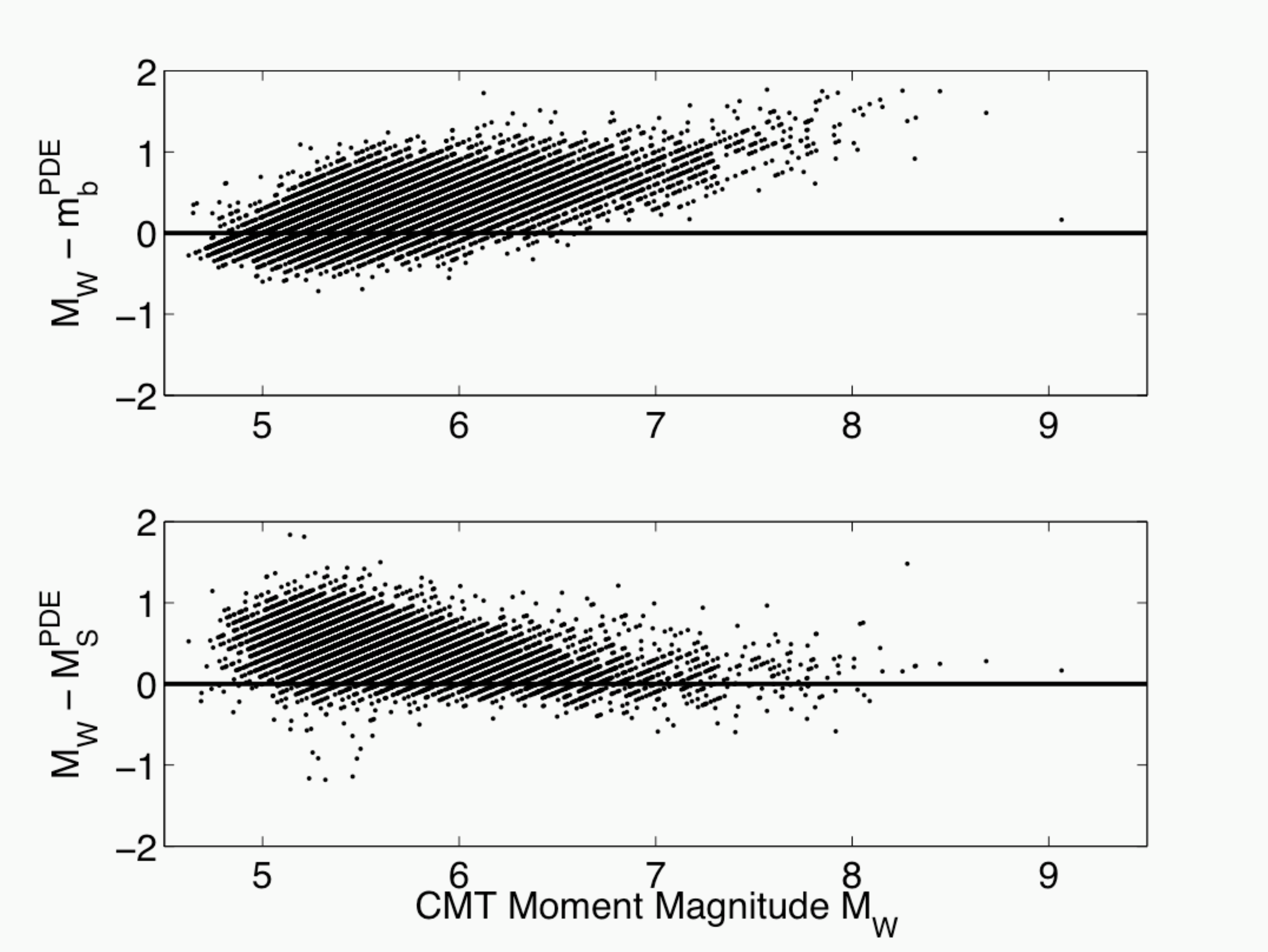}
\end{center}
\caption{\label{Mwmbms} Estimating {\it inter}-magnitude uncertainties by comparing the CMT moment magnitude $M_W$ with its corresponding body wave magnitude $m_b$ (a) and surface wave magnitude $M_S$ (b) from the PDE catalog. }
\end{figure}

We used the global Harvard CMT catalog from 1 January 1976 until 31 December 2005 (available from http://www.globalcmt.org/CMTfiles.html in gzipped ``ndk" format). We found $24583$ events listed. Of these, we selected all events that were assigned the source ``PDE" ($21450$ events) and converted their scalar seismic moments to moment magnitudes using the equation $M_W= 2/3 \ \log_{10} (M_0)-6$ [{\it Kagan}, 2003]. We found $21435$ $m_b$ and $13363$ $M_S$ values which we subtracted from their corresponding $M_W$ magnitudes. Figure \ref{Mwmbms} shows the resulting differences as a function of the Harvard CMT $M_W$. There are systematic trends and random scatter. The body-wave magnitude $m_b$ systematically underpredicts $M_W$ for about $M_W>5.2$. Since $m_b$ is based on relatively short periods, the energy in these frequency bands does not increase beyond this value and the scale saturates. The S-wave magnitude $M_S$, on the other hand, underpredicts $M_W$ systematically but especially for $M_W<7$. 

\begin{figure}[!ht]
\begin{center}
\includegraphics[width=16cm]{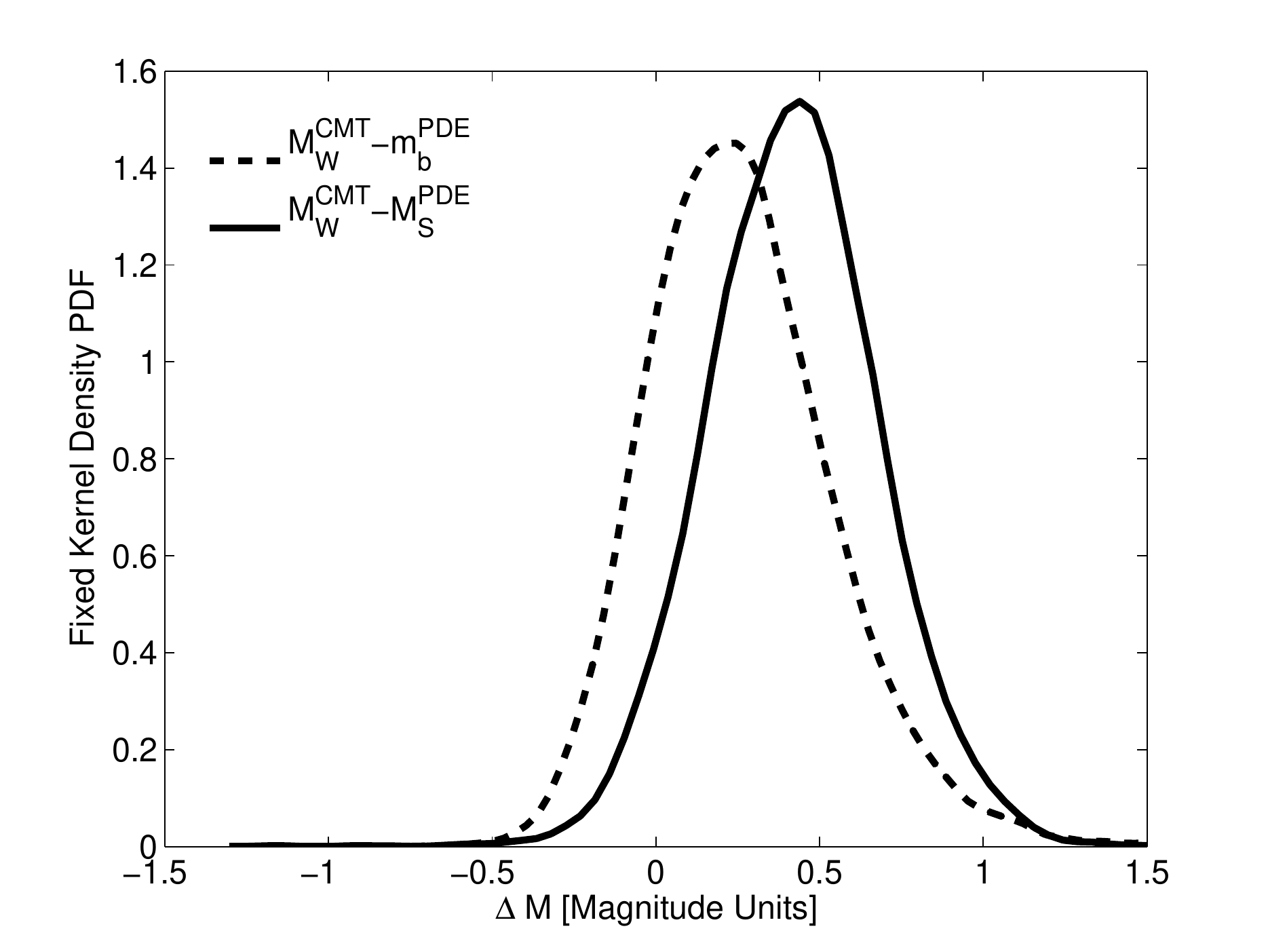}
\end{center}
\caption{\label{Mwmbmsker} Kernel density estimates of the probability density functions of the differences between the Harvard CMT moment magnitudes $M_W$ and their corresponding body wave $m_b$ (dashed) and surface wave $M_S$ (solid) magnitude estimates from the PDE. The means of the data are 0.26 for $m_b$ and 0.42 for $M_S$. The standard deviations are about 0.29 for $m_b$ and 0.26 for $M_S$. 
}
\end{figure}

Figure \ref{Mwmbmsker} shows fixed kernel density estimates of the probability density functions (pdfs) of the two sets of differences. The systematic shifts of both pdfs indicate systematic under-estimation of $M_W$. The means of the data are 0.26 for $m_b$ and 0.42 for $M_S$. The widths of the pdfs indicate the random scatter. The standard deviations are approximately 0.29 for $m_b$ and 0.26 for $M_S$. 

For context, an ETAS model would predict 10 times as many aftershocks if the magnitude unit were inflated spuriously by one magnitude unit! These differences can have a profound impact on global testing experiments. 

\subsubsection{Duration Magnitude vs. Local Magnitude in the NCSN Catalog}
\label{inter2}

The NCSN catalog reports both coda duration magnitude $M_D$ and maximum amplitude (local) magnitude $M_L$ in its Hypoinverse output format, available from  \url{http://www.ncedc.org/ncedc/catalog-search.html}. We used data from 1 January 1984 until 31 December 2006 in the region 33$^o$ to 43$^o$ latitude and -120$^o$ to -115$^o$ longitude. We selected earthquakes larger than the magnitude threshold $m_{th}=3$, leaving a total of 4679 events. We found 4595 reported MAD values for duration magnitudes $M_d$ and 2711 MAD values reported for local magnitudes $M_L$. 

Whenever both magnitudes were available for the same event, we compared their values. However, we additionally required that at least one of the two magnitudes be equal to or larger than $M_{(\cdot)}=3$. Despite selecting a magnitude cut-off $m_{th}=3$ in the search algorithm on the website \url{http://www.ncedc.org/ncedc/catalog-search.html}, we found 74 events out of the total 4679 where both $M_D$ and $M_L$ were smaller than the prescribed cut-off. The extreme case was an event with $M_D=0.35$ and $M_L=-0.43$ (we presume that the cut-off magnitude corresponds to the magnitude reported to the ANSS, which may be different, see beginning of section \ref{NCSNnoise}). Removing these 74 events, we are left with 4605 events which obey the condition that at least one of the two magnitudes be equal to or larger than 3. Out of these events, we found 2733 events for which both $M_D$ and $M_L$ were reported. We then calculated the difference $\Delta=M_D-M_L$ for these 2733 events. 

\begin{figure}[!ht]
\begin{center}
\includegraphics[width=16cm]{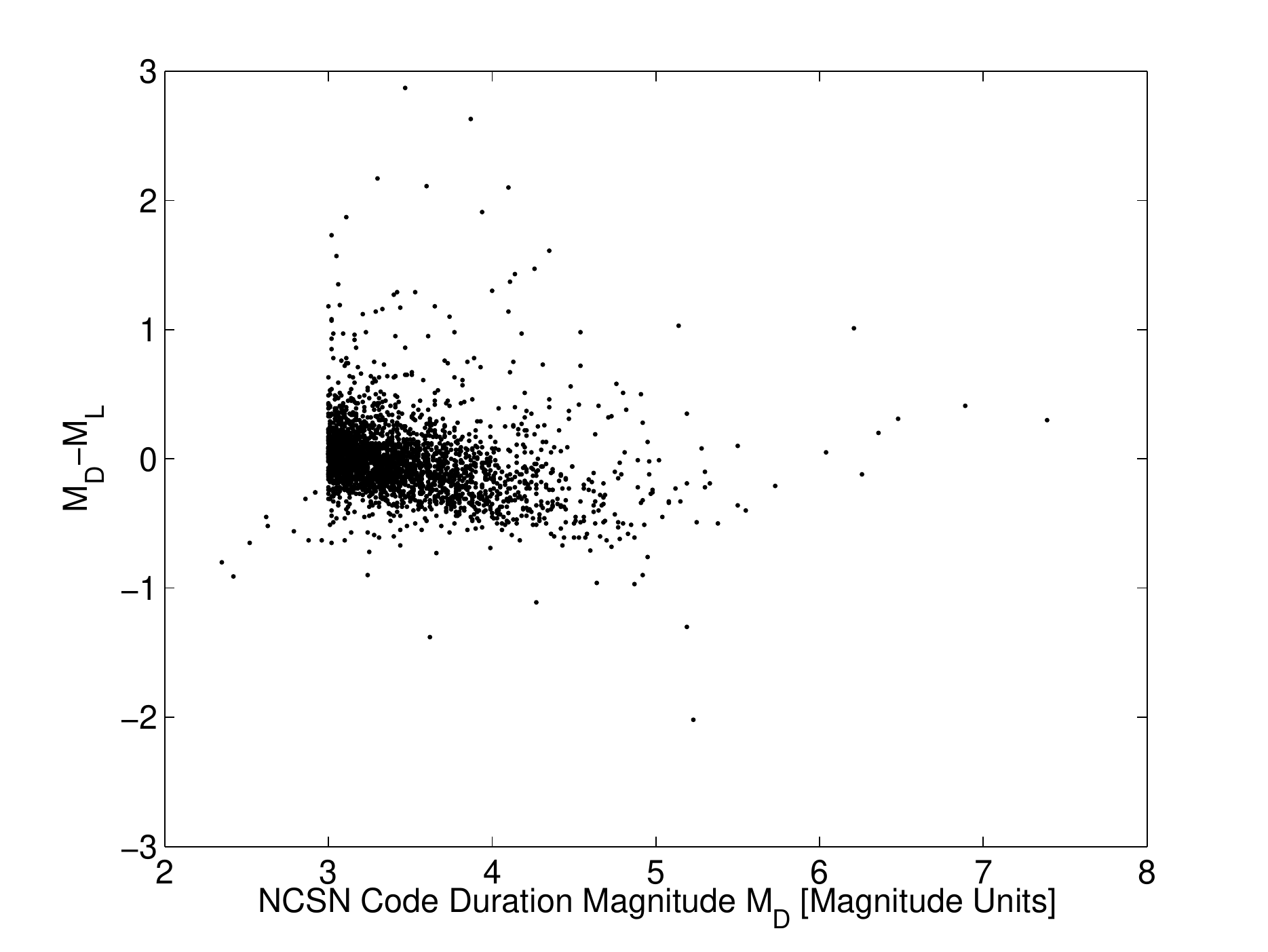}
\end{center}
\caption{\label{Deltamd} Differences between the coda duration magnitude $M_D$ and the local amplitude magnitude $M_L$ versus $M_D$ in the NCSN catalog whenever at least one is larger than $3$.
}
\end{figure}

Figure \ref{Deltamd} shows the differences as a function of $M_D$. Recall that at least one of the two magnitudes must be larger than 3 (but not necessarily both). Few events are reported with $M_L>3$ and $M_D<3$. On the other hand, many events are reported with $M_L<3$ and $M_D>3$. It is hard to discern a trend visually, but $M_L$ seems to under-predict $M_D$ up to about $M_D=3.5$, then over-predict until about $M_D=5.5$. 

\begin{figure}[!ht]
\begin{center}
\includegraphics[width=16cm]{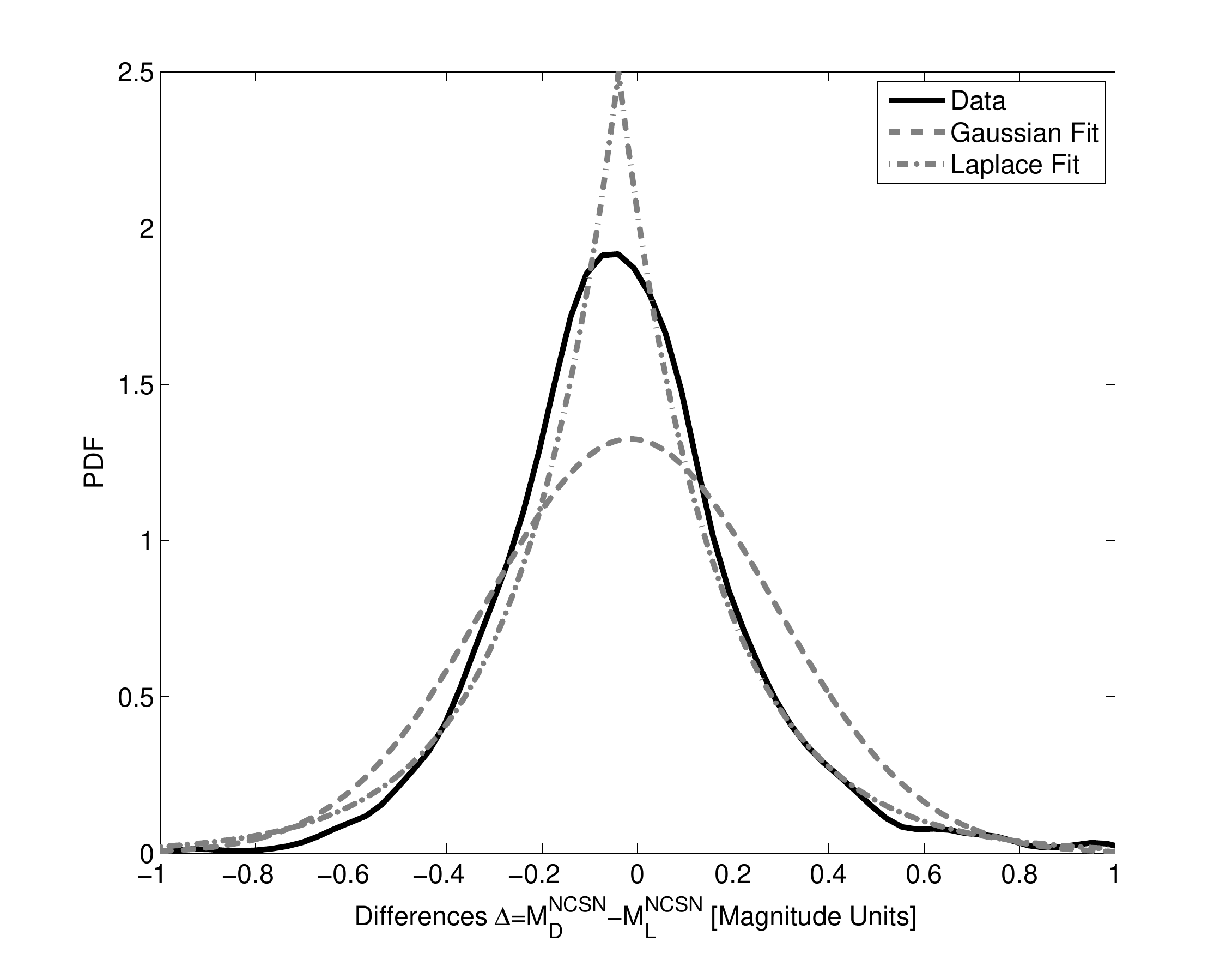}
\end{center}
\caption{\label{Deltafit} Kernel density estimate (solid) of the probability density function of the differences $\Delta$ between the duration magnitude $M_D$ and the local magnitude $M_L$ reported by the NCSN. Also shown are a Gaussian fit with mean $-0.0153$ and standard deviation $0.3$ (dashed); and a fit to a Laplace pdf with median $-0.04$ and e-folding scale parameter $0.2$ (dotted). 
}
\end{figure}

Figure \ref{Deltafit} shows a fixed kernel density estimate (solid) of the probability density function of the differences $\Delta$. The largest difference between the two was 2.87 magnitude units (note that the x-axis was cut at $\pm 1$). The mean of their differences was $-0.015$, essentially showing no systematic bias. The standard deviation was $0.3$, while the e-folding scale parameter is 0.2. Also shown are fits of the data to a Gaussian distribution (dashed; mean equal to $-0.015$ and standard deviation equal to $0.3$) and to a Laplace distribution (dash-dotted; median equal to $-0.04$ and scale parameter equal to 0.2). 
While neither fit approximates well the central part of the pdf, the Laplace distribution provides much better fits to the bulk and tails of the data. The Gaussian distribution significantly underestimates the probability of outliers. 

\subsection{Summary of Magnitude Uncertainty}
Firstly, we compared estimates of the moment magnitude from the CMT and USGS MT catalogs. We found that the Laplace distribution is a good approximation to the bulk of the magnitude differences but it underestimates the tails. Our characterization of the entire distribution of the magnitude differences implies that individual uncertainties are distributed with tails that decay exponentially or even more slowly. 

Secondly, we analyzed a data set directly relevant to CSEP predictability experiments. We analyzed MAD values, a magnitude uncertainty measure, reported in the NCSN's authoritative region in the ANSS. We found (i) MAD values fluctuate up to $1.71$ with an average of $0.15$,  (ii) there is no evidence that magnitude uncertainty decreases with increased magnitude, (iii) the region may not be complete down to $m_d=3$, (iv) MAD values less than $0.1$ seem unreliable, and (v) the 99th percentile of MAD values is only reached at $0.59$. 

We also considered inter-magnitude uncertainties. These can be extremely large and include systematic differences. We found that PDE body and surface wave magnitudes are systematically smaller (with mean 0.26 and 0.42, respectively) and randomly scattered (with standard deviations 0.29 and 0.26, respectively). 

Finally, we studied the differences between NCSN's duration and local magnitudes. We found that the Laplace distribution again provided an adequate fit to the differences with scale factor $0.2$ so that individual uncertainties have exponential or fatter-than-exponential tails. 

\section{Impact of Magnitude Noise on Seismic Rate Estimates}
\label{impact}

In the previous section, we studied magnitude uncertainties. How do these magnitude uncertainties propagate to seismic rate estimates and to forecasts? How can they influence forecast evaluation tests and the rate of type I and II errors? In particular, how could they influence the CSEP predictability experiments? In this section, we address the first question of the impact of magnitude noise on the estimates of seismic rates in short term clustering models. We use the knowledge of magnitude uncertainties we have gained in the previous section to model to simulate magnitude noise. In section \ref{LT}, we conduct a numerical experiment to begin answering the second and third question. 

\subsection{A Simple Aftershock Clustering Model}

Most short term seismicity models use three statistical laws to extrapolate rates into the future. These are the Gutenberg-Richter law for magnitude-frequency statistics, the Omori-Utsu law for the temporal distribution of aftershocks and the productivity law for the expected number of offspring of a mother-shock. Models based on these laws include the Epidemic Type Aftershock Sequence (ETAS) Model [{\it Ogata}, 1988], Short Term Earthquake Probabilities (STEP) [{\it Gerstenberger et al.}, 2005], the model by {\it Reasenberg and Jones} (1989, 1994), the Epidemic Rate and State (ERS) model by {\it Console et al.} (2007) and the model by {\it Kagan and Knopoff} (1987). Although there are important differences between these models, all of them employ the above-mentioned three laws to forecast events. In particular, all of them use the so-called productivity law, defined below in equation (\ref{rho}), in which the number of expected aftershocks is an exponential function of  the mother magnitude. 

We choose the Poisson cluster model as the basis of our analysis [see {\it Daley and Vere-Jones}, 2003]. It is simpler than the above models, yet it captures the essence of the clustering phenomenon through the above three laws. In particular, it preserves the exponential sensitivity of the number of expected aftershocks on the magnitude of the mother-shock. We expect that the conclusions obtained below concerning the impact the magnitude errors on the uncertainty of predicted seismic rates carry over to the above mentioned models. For models that include secondary triggering, the impact of magnitude noise may even be strongly amplified. 

The model can be described as follows. It consists of two processes: a cluster center process and an aftershock process. The cluster center process creates the mother events (also called main shocks or background) which have not been triggered. This process is Poissonian and governed by a homogeneous rate $\lambda_c$. The magnitudes (or marks) of the cluster centers are drawn independently from the Gutenberg-Richter distribution [{\it Gutenberg and Richter}, 1954]: 
\be
p_m(m) = \beta e^{-\beta (m-m_d)}, \ \ \ m \geq m_d
\label{GR}
\ee
where $m_d$ is an arbitrary cut-off determined by the detection threshold and $\beta = b\log(10)$ is a constant. We denote the marked cluster center process by $\{t_{i_c}, m_{i_c}  \}_{1\leq i_c \leq N_c}$. Each of these mothers can trigger direct aftershocks. In contrast to cascading models such as ETAS, there is no secondary triggering, allowing for a simplified analytical treatment and faster simulations. The number of aftershocks is a random number drawn from a Poisson distribution with expectation given by the productivity law: 
\be
\rho(m_{i_c}) = k \ e^{a(m_{i_c}-m_d)} 
\label{rho}
\ee
where $k$ and $a$ are constants and $m_{i_c}$ are the magnitudes of the mothers. The productivity law captures the exponential dependence of the number of aftershocks on the magnitude of the mother. This exponential dependence suggests that fluctuations in the magnitudes due to noise will strongly affect any forecast, as we are going to demonstrate analytically and by numerical simulations. 

The threshold $m_d$, which measures the size of the smallest triggering earthquake below which earthquakes do not trigger, is arbitrarily set to the detection threshold. This unjustified assumption is known to bias the clustering parameters  [{\it Sornette and Werner}, 2005a, 2005b].

The triggered events are distributed in time according to the Omori-Utsu law [{\it Utsu et al.}, 1995]:
\be
\Phi(t-t_{i_c}) = { 1 \over (t-t_{i_c}+c)^p}
\label{Omori}
\ee
where $c$ and $p$ are constants and the $t_{i_c}$ are the occurrence times of the cluster centers. The marks of the aftershocks are distributed in the same manner as their mothers according to the Gutenberg-Richter law (\ref{GR}).

In summary, the rate of events (including aftershocks) of the marked Poisson cluster process is completely defined by its conditional intensity (or rate) [see {\it Daley and Vere-Jones}, 2003]: 
\be
\lambda (t,m | H^c_t, \theta)= p_m(m) \left(\lambda_c + \sum_{i_c | t_{i_c}<t} {k\ e^{a(m_{i_c}-m_d)}\over (t-t_{i_c}+c)^p} \right)
\label{intensitymag}
\ee
where $H^c_t$ is the history up to time $t$ which need only include information about the cluster centers $\{t_{i_c}, m_{i_c}  \}_{1\leq i_c \leq N_c}$, as aftershocks do not trigger their own aftershocks and do not influence the future. The set of parameters $\theta=\{\beta,\lambda_c, k, a, m_d, c, p \}$ are assumed to be known. 

The (unmarked) intensity above the detection threshold $m_d$ is simply given by integrating over $m$: 
\be
\lambda (t | H^c_t, \theta)= \lambda_c + \sum_{i_c | t_{i_c}<t} {k\ e^{a(m_{i_c}-m_d)}\over (t-t_{i_c}+c)^p}
\label{intensity}
\ee

In section \ref{LT}, we will use a spatio-temporal Poisson cluster process to mimic the CSEP experiment and its evaluation as closely as possible. The model is defined by the conditional intensity at time $t$ and location ${\bf r}$: 
$$
\lambda (t,{\bf r},m | H^c_t)= p_m(m) \times
$$
\be \left(\lambda_c + \sum_{i_c | t_{i_c}<t} {k\ e^{a(m_{i_c}-m_d)}\over (t-t_{i_c}+c)^p }   { \mu \ (0.02 \cdot  10^{0.5 m_{i_c}})^{\mu} \over  (0.02  \cdot 10^{0.5 m_{i_c}} + |{\bf r} - {\bf r_{i_c}}|)^{1+\mu}  }
\label{lambdaspt} \right)
\ee
where we added a commonly used spatial decay function [see e.g. {\it Helmstetter et al.,} 2005].

\subsection{Magnitude Noise}

To study the impact of magnitude noise on seismic rate estimates and forecasts, we assume that each (true) magnitude is perturbed identically and independently by an additive noise term $\epsilon$
\be
m^o_{i_c}=m^t_{i_c} + \epsilon_{i_c}~.
\label{magnoise}
\ee
We do not need to perturb the magnitudes of the aftershocks because they have no influence on the future seismicity rate. 

We use our results on magnitude uncertainties from section \ref{dm} by modeling noise according to a Laplace distribution characterized by zero mean (unbiased) and a scale (e-folding) parameter $\nu_c$:
\be
p_{\epsilon} (\epsilon ) = { 1 \over 2 \nu_c} e^{\left(  - { |\epsilon | \over \nu_c}   \right)}
\label{noise2}
\ee
We believe that this is a conservative estimate of the distribution of the magnitude noise because the Laplace distribution under-estimates the occurrence of large error outliers. In general, therefore, the fluctuations of the seismic rate are likely to be even larger than calculated below. 

We assume the parameters are known. In particular, we assume below that the seismic rate estimates from noisy magnitudes use the same parameters as the ``true" rate. We do so to isolate the effect of magnitude uncertainties. A comprehensive analysis of uncertainties including trade-offs between parameter and data uncertainty is beyond the scope of this article and is most likely extremely model-dependent.

\subsection{Fluctuations in the Seismic Rates Due to Noisy Magnitudes}

To investigate the fluctuations of the seismic rate estimates due to noisy magnitudes, we compare the seismic rates estimated from perturbed magnitudes with the true seismic rate. The perturbed rates from the observable, noisy magnitudes $m^o_i$ are given by:
\be
\lambda^p (t | H^c_t)= \lambda_c + \sum_{i_c | t_{i_c}<t} {k\ e^{a(m^o_{i_c}-m_d)}\over (t-t_{i_c}+c)^p}
\label{lambda_p}
\ee
We consider the deviation $\Delta \lambda(t)$ of the perturbed rates from the true rate, given by: 
\be
\Delta \lambda (t|H^c_t)   = \lambda^p (t | H^c_t) - \lambda (t | H^c_t)
\label{deltalambda}
\ee
In the remainder of this section, we characterize the fluctuations $p(\Delta \lambda)$ of the deviations from the true rate $\Delta \lambda (t|H^c_t)$. In particular, we would like to know how strong the deviations can be for a given catalog and whether one can make general statements for any catalog. 

\begin{figure}[tbp]
\begin{center}
\includegraphics[width=16cm]{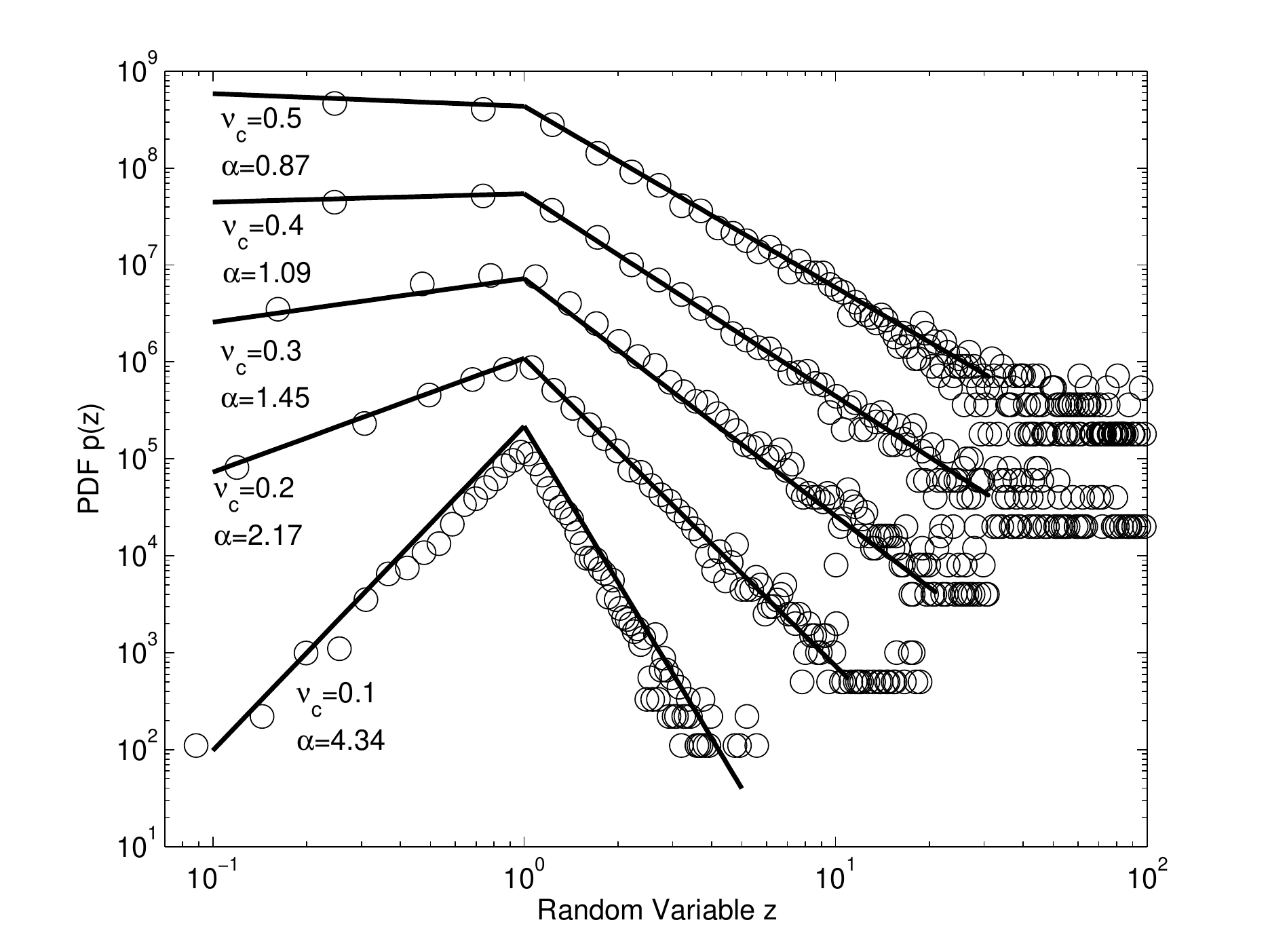}
\end{center}
\caption{\label{z} Theoretical and simulated probability density function (pdf) of the random variable $z=(\exp(a \cdot \epsilon)-1)$ for various choices of the scale parameter of the noise $\nu_c$ and assuming $a=2.3$. The curves are shifted for clarity. 
}
\end{figure}

To obtain some properties of the distribution of $\Delta \lambda (t|H_t)$, we rewrite equation (\ref{deltalambda}) as a finite sum over the product $w_i \cdot z_i$ (see Appendix \ref{A1}):
 \be
\Delta \lambda (t|H^c_t)   =\sum_{i_c | t_{i_c}<t} w_i z_i
\label{dl}
\ee
where 
\be
w_i={k e^{a(m^t_i-m_d)} \over (t-t_i+c)^{p}}
\label{weights}
\ee
are quenched weights that depend on the true magnitudes and occurrence times of the cluster centers, while 
\be
z_i=e^{a\epsilon_i}-1
\label{zi}
\ee
are random variables due to the random noise $\epsilon$. From equation (\ref{pdfz}) in Appendix \ref{A2}, we see that the $z_i$ are power-law distributed with an exponent $\alpha= 1/\nu_c a$ that depends inversely on the size of the noise $\nu_c$ and the exponent of the productivity law $a$. In Figure \ref{z}, we show the theoretical and simulated pdf of the random variable $z$ for several values of the noise level $\nu_c$. 

Equation (\ref{dl}) together with the knowledge that the random variables $z_i$'s are power law distributed implies that the following proposition is true: 

\vspace{0.5cm}

\noindent \textit{Proposition 1:}  
The deviations $\Delta \lambda (t|H^c_t) $ of the perturbed seismic rates from the true seismic rate due to magnitude noise converge to bounded random variables with a distribution having a
power law tail  for $\Delta \lambda \to \infty$:
\be
p(\Delta \lambda) \sim {C_{ \Delta \lambda} \over (\Delta \lambda)^{1+\alpha}}  \hspace{1cm}  \rm{with} \hspace{0.3cm} \alpha={1\over \nu_c a}
\label{prop1}
\ee
\vspace{0.3cm}
\noindent 
and with a scale factor given by
\be
C_{\Delta \lambda}=\sum_{i=1}^{N_c} w_i^\alpha~,
\label{jghjblw}
\ee
where the sum is over the $N_c$ earthquakes in the catalog that occurred before
the present time $t$ and the $w_i$'s are given by (\ref{weights}).

\vspace{0.5cm}

\noindent \textit{Proof:} See Appendix \ref{A3}. 

\vspace{0.5cm}

\noindent \textit{Remarks:} 
\begin{enumerate}

\item The deviations of a rate estimate based on noisy magnitudes from the exact rate are power-law distributed in the tail. Estimates of a seismic rate may therefore be wildly different from the ``true" rate, simply because of magnitude uncertainties.

\item The exponent $\alpha$ determines how broadly the seismic rate estimates are distributed around the ``true" rate. The exponent sensitively depends on the noise level $\nu_c$ and the productivity exponent $a$. For the often quoted value $a=\ln(10)=2.3$ [e.g. {\it Felzer et al.}, 2004] and for $\nu_c=(0.1,0.2,0.3,0.5)$, one obtains $\alpha=(4.34, 2.17, 1.45, 0.87)$, respectively. Even for relatively small levels of noise $\nu_c \ge 0.22$, the variance of $\Delta \lambda$ does not exist ($\alpha \leq 2$), while for $\nu_c \ge 0.44$, the average of $\Delta \lambda$ does not exist ($\alpha \leq 1$)! 

\item The power law tail does not depend on the specific history of the magnitudes and occurrence times. The same power law tail holds for any catalog. On the other hand, the scale factor $C_{\Delta \lambda}$ depends sensitively on the specific realization via the magnitudes and occurrence times (see Proposition 2). 

\item Short term earthquake forecasts are directly (sometimes indirectly) taken from such seismic rate estimates. The accuracy of the forecasts cannot be better than the seismic rate estimates. 

\end{enumerate}

\begin{figure}[tbp]
\begin{center}
\includegraphics[width=16cm]{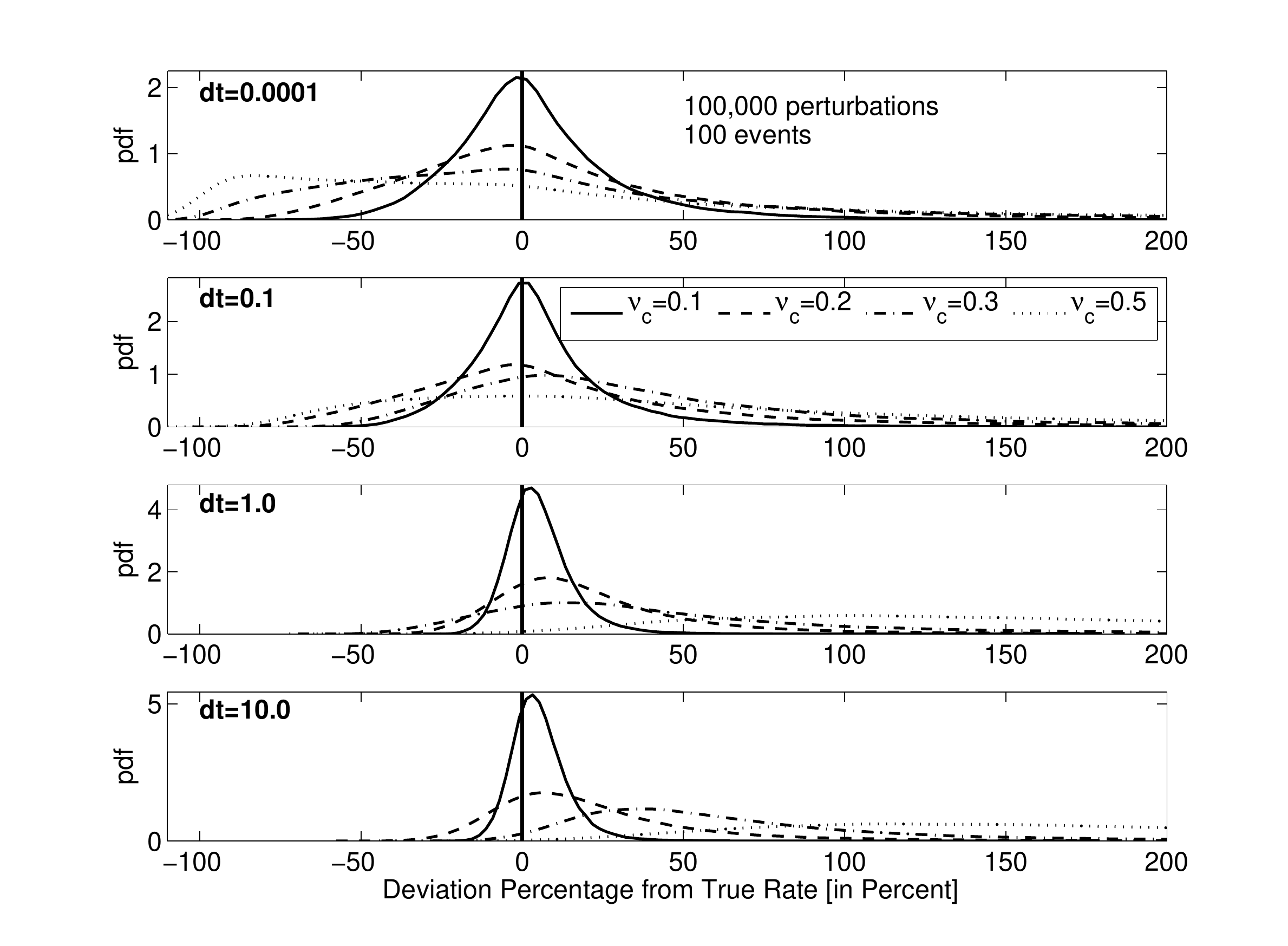}
\end{center}
\caption{\label{KDE} Simulated probability density functions of the ``perturbed" seismic rates due to noisy magnitudes shown as a percent deviation from the true rate for noise levels $\nu_c=(0.1, 0.2, 0.3, 0.5)$. We calculate seismic rates at a time lag $dt$ after the 100th cluster center event. From top to bottom, $dt=(0.0001, 0.1, 1, 10)$. Seismic rate estimates (and hence forecasts) are wildly distributed around the true value (solid vertical lines). 
}
\end{figure}

As a demonstration of the strong fluctuations, Figure \ref{KDE} shows simulations of the seismic rate deviations for various levels of noise $\nu_c=(0.1, 0.2, 0.3, 0.5)$. In each case, we simulate $N_c=100$ cluster centers according to a Poisson process and randomly draw their magnitudes from the GR law. We then choose a time lag $dt$ (days) between the last event in the cluster center process and the time at which we want to evaluate (forecast) the seismic rate. We calculate the ``true" seismic rate from equation (\ref{intensity}), using the parameter set $\theta=\{\beta=2.3,\lambda_c=1, k=0.01, a=2.3, m_d=3, c=0.001, p=1.2\}$ and using the ``true" simulated magnitudes. Next, we generate ``perturbed" catalogs by adding noise to the simulated magnitudes $100,000$ times according to the double exponential noise defined in (\ref{noise}). For each perturbed catalog, we recalculate the ``perturbed" seismic rate as before using equation (\ref{intensity}) but replacing the ``true" magnitudes by noisy magnitudes.  

\begin{figure}[tbp]
\begin{center}
\includegraphics[width=16cm]{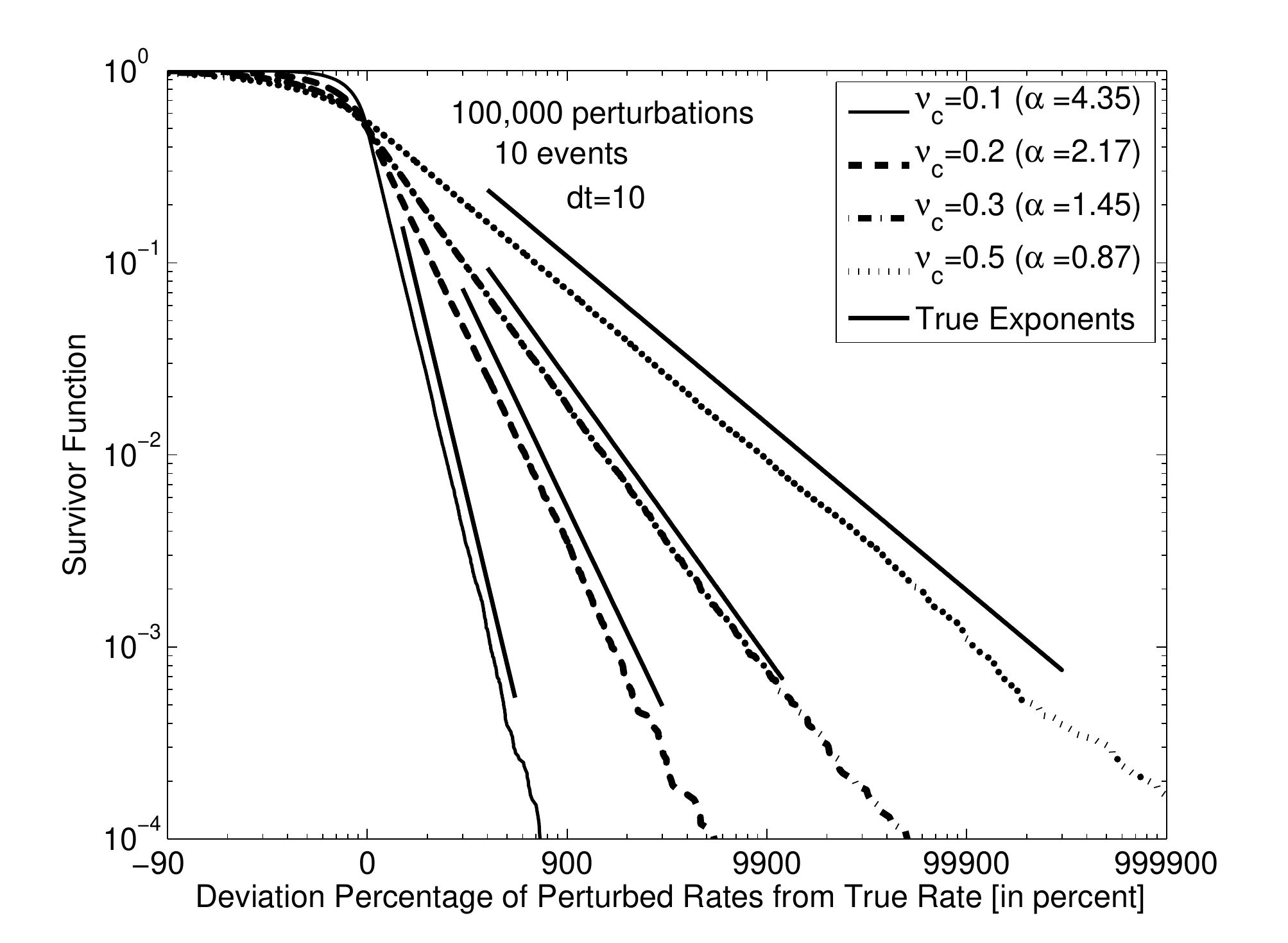}
\end{center}
\caption{\label{logCDF1} Survivor functions of the simulated ``perturbed" seismic rate estimates (and forecasts) shown as a deviation in percent from the simulated ``true" rate for noise levels $\nu_c=(0.1, 0.2, 0.3, 0.5)$ and time lag since the last event in the process $dt=0.0001$. The survivor functions 
are approximately parallel to the straight lines which are guides to the eye with theoretically predicted exponents given by $\alpha=(4.34, 2.17, 1.45, 0.87)$, respectively. Even for small noise level $\nu_c=0.2$, 10 percent of the rate estimates over-predict the rate by 100 percent! 
}
\end{figure}

\begin{figure}[tbp]
\begin{center}
\includegraphics[width=16cm]{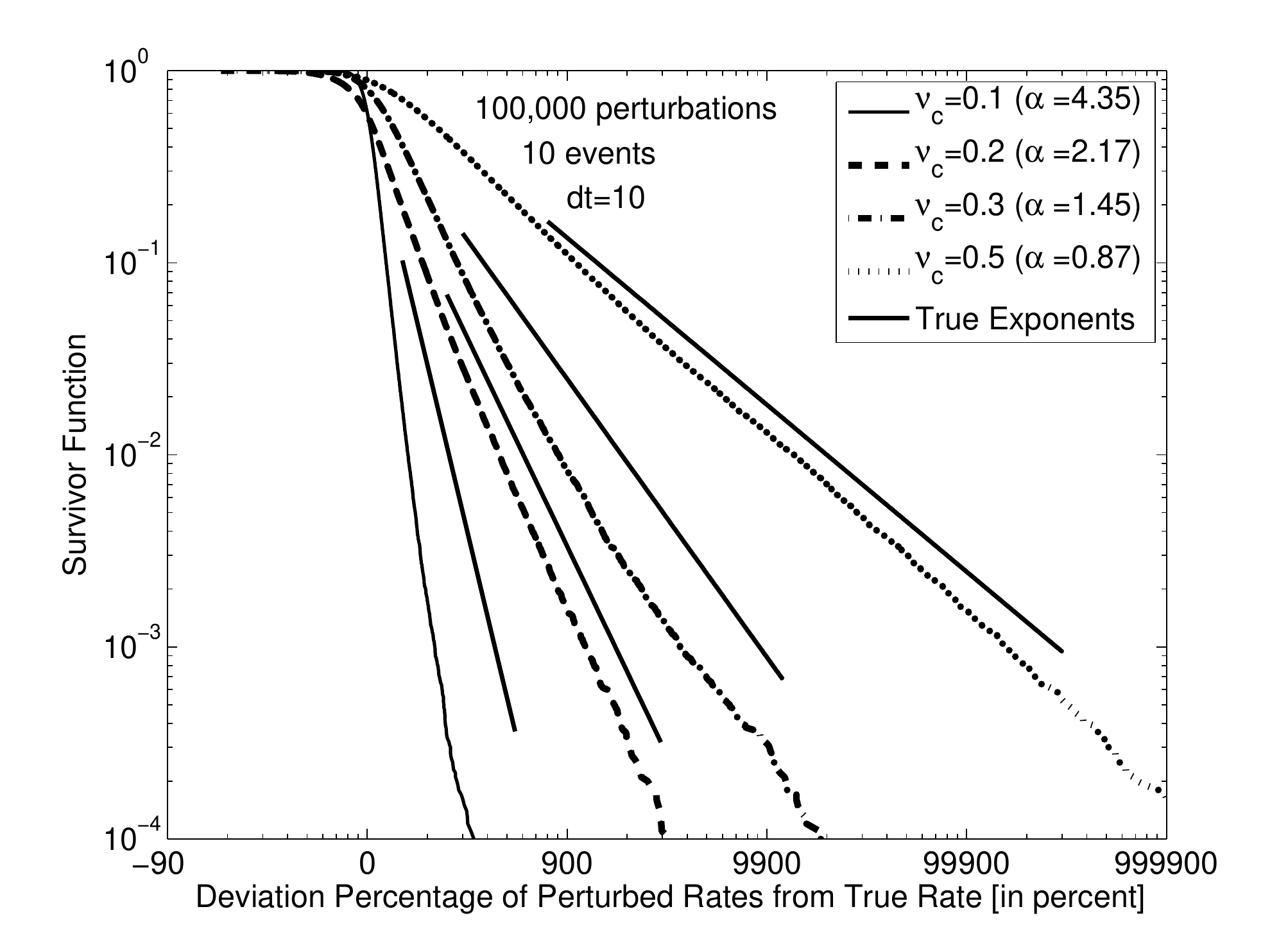}
\end{center}
\caption{\label{logCDF2} Same as Figure \ref{logCDF1} except that perturbed and true rates are evaluated at $dt=10$: log-log plots of the survivor functions of the simulated ``perturbed" seismic rate estimates (and forecasts) shown as a deviation in percent from the simulated ``true" rate for noise levels $\nu_c=(0.1, 0.2, 0.3, 0.5)$. The slopes of the straight lines are given by the asymptotic predicted exponents  $\alpha=(4.34, 2.17, 1.45, 0.87)$, respectively.  
}
\end{figure}

Figure \ref{KDE} shows fixed kernel density estimates of the normalized differences between the perturbed and the true rates. In each panel, the deviations are shown for four different noise levels $\nu_c=(0.1, 0.2, 0.3, 0.5)$. The different panels (top to bottom) are different choices of the time lag $dt=\{0.0001, 0.1, 1, 10 \}$ days since the last event in the process at which the rates are calculated. Figures \ref{logCDF1} and \ref{logCDF2} show double logarithmic plots of the survivor functions for the two extreme cases $dt=0.0001$ and $dt=10$, respectively. 

The seismic rate estimates are very broadly distributed, but with different dependence
on the specific catalog, depending on the time horizon $dt$. 
For small $dt$ (smaller than the average inter-event time $1/\lambda_c$ between successive main shocks), the only relevant event that determines the distribution of the perturbed rates is the last one. Therefore, the (normalized) distributions for small $dt$ are almost identical to the distribution of the random variable $z$ (up to the weight pre-factor associated with the last event). However, as $dt$ increases to become comparable to and larger than $1/\lambda_c$, the rate is no longer dominated solely by the last event. Previous events can become increasingly significant in determining the sum compared to the last event. Consider two earthquakes of similar magnitudes occurring at $t_1=0$ and $t_2=10$ respectively. At time $t=11$, $t-t_2=1$ and $t-t_1=11$ so that earthquake 2 will dominate at early times by the effect of the Omori law on the weights, since $w_2/w_1 (t=11) = (11+c)^p / (1+c)^p \sim 11$. But at $t=20$, say, the weight $w_1$ of the first earthquake has decreased only by a factor $2^p$ while the weight $w_2$ of the second earthquake has decreased by a factor $10^p$, so that the ratio is now $w_2/w_1 (t=20) = (20+c)^p / (10+c)^p \sim 2$. Since the weights are so strongly stochastic, each particular catalog realization will have a different number of events with vastly different weights that are ``felt" by the rate as $dt$ increases. 

We give some concrete numbers for illustration for two cases, $dt=0.0001$ (Figure \ref{logCDF1}) and $dt=10$ (Figure \ref{logCDF2}).
With our parameter choices $c=0.001$ and $\lambda_c=1$, this ensures that 
$dt=0.0001$ corresponds to the regime dominated by the last event in the catalog 
while the case $dt=10$ will show the impact of many past events.
For $dt=0.0001$ and realistic $\nu_c=0.2$, 80$\%$ of the seismic rate estimates (and hence forecasts) deviate by more than $10\%$; almost two thirds are off by more than $20\%$; and almost one third are off by more than $50\%$. Even fluctuations of $100\%$ occur $11\%$ of the time. For larger levels of noise, much stronger fluctuations can occur. For instance, $28\%$ of rate estimates are off by more than $100\%$ for $\nu_c=0.5$.
For the case where seismic rate estimates (forecasts) are made for $dt=10$ after the last event, the percentages depend strongly on the particular cluster center realization. 

Apart from the dependence on a particular realization, there is another consequence of the increasing importance of previous events with time since the last event. As previous events are summed over, the distribution of the sum tends towards its stable law (either Gaussian or L\'evy). For $\alpha>2$ ($\nu_c<0.22$), the central limit theorem applies to the body of the distribution (though not the tails, see e.g. {\it Sornette}, 2004) and tends to organize the distribution of the deviations towards a Gaussian. At the same time, only a finite number of terms effectively control the rate so that there is no asymptotic convergence. Nevertheless, there may be a tendency for the body of the distribution to become Gaussian. For $0<\alpha<2$ ($\nu_c>0.22$), in contrast, the distribution will tend towards a L\'evy law with a power law tail exponent equal to $\alpha$, keeping intact the original power law tail. 

This discussion can be summarized by the following proposition, which emphasizes that
the scale factor $C_{\Delta \lambda}$ is ``non-self-averaging'' [{\it M\'ezard et al.}, 1987].
\vspace{0.5cm}

\noindent \textit{Proposition 2:}  
\begin{enumerate}
\item For $p \alpha = {p \over \nu_c a}  > 1$,
the scale factor $C_{\Delta \lambda}=\sum_i^{N_c} w_i^\alpha$ given in \textit{Proposition 1}
(or the typical scale $C_{\Delta \lambda}^{1 /\alpha}$) of the distribution of the deviations of the perturbed rates from the true rate converges to a finite value as $N_c \to \infty$. 
\item For  $p \alpha = {p \over \nu_c a}  \leq 1$, the scale factor $C_{\Delta \lambda}=\sum_i^{N_c} w_i^\alpha$ diverges $\to \infty$ as $N_c \to \infty$. This means that the deviations $\Delta \lambda (t|H_t)$ of the perturbed rates from the true rate diverge as the duration of the 
catalog used to calculate them increases without bound. 
\item In the regime  $p\alpha > 1$ for which $C_{\Delta \lambda}$ converges almost surely
to a finite value as $N_c \to \infty$,  $C_{\Delta \lambda}$ remains a random variable
sensitively dependent on the specific catalog, $C_{\Delta \lambda}$ being
distributed according to a non-degenerate distribution.  In symbols, 
\be
C_{\Delta \lambda} \ \nrightarrow \  \langle C_{\Delta \lambda} \rangle \hspace{1cm} \rm{as} \hspace{1cm} N_c \to \infty~,
\label{prop2}
\ee
where $ \langle C_{\Delta \lambda} \rangle$ denotes the average of $C_{\Delta \lambda}$
over many different catalogs.
This means that the value $C_{\Delta \lambda}$ is catalog specific and changes from catalog to catalog  and from realization to realization. This property is known as ``lack of self-averaging'' [{\it M\'ezard et al.}, 1987]. 
\end{enumerate}

\vspace{0.5cm}

\noindent \textit{Proof:} See Appendix \ref{A4}. 

\vspace{0.5cm}
The divergence of the scale factor $C_{\Delta \lambda}$ and of the deviations $\Delta \lambda (t|H_t)$
described by item 2 of Proposition 2
occurs only for rather large magnitude errors. For instance, for 
$p=1.2$ and $a=2.3$, the noise level needs to be larger than $\nu_c=0.52$ for $p \alpha = {p \over \nu_c a}  \leq 1$ to hold. According to our previous analysis in Section 2, we can expect $\nu_c$
to lie in the range $0.1-0.3$ typically, so that the regime $p \alpha = {p \over \nu_c a}  > 1$ in Proposition 2 is likely to be the most relevant one.

In summary, for large $dt$, there is a trade-off between some averaging for small noise levels (i.e. a tendency towards a Gaussian shape), the dependence on the specific catalog realization (in particular, the last few events) and the power law tail proven in Proposition 1. These effects can be seen by comparing the survivor functions in Figure \ref{logCDF1}, which for $dt=0.001$ are pure power laws for $\Delta \lambda>0$, with Figure \ref{logCDF2} for $dt=10$. For $\nu_c=0.3$ and $\nu_c=0.5$ (L\'evy regime), the power law remains largely intact even for the body of the distribution. For $\nu_c=0.1$ and $\nu_c=0.2$, the body is not well approximated by the power law, while the tails still show significantly larger outliers than expected from a Gaussian. 

We now return to the explicit values and percentages of the deviations for the case $dt=10$. For realistic $\nu_c=0.2$, about $70\%$ are off by more than $10\%$; $45\%-55\%$ deviate by more than $20\%$; roughly $20\%$ of seismic rate estimates over-predict the true value by more than $50\%$. 

In contrast to the noise levels for which $\alpha>2$, the fluctuations actually increase for $\nu_c=0.3$ and $\nu_c=0.5$ compared to the case $dt=0.0001$. For $\nu_c=0.5$, $90\% - 95\%$ are off by more than $10\%$; $80\% - 90\%$ are off by more than $20\%$; $70\% - 75\%$ are off by more than $50\%$; and roughly one half are off by more than $100\%$!

To conclude this section, we restate the main results: (i) Seismic rates estimated from noisy data deviate strongly from their ``true" value; (ii) If magnitude uncertainties are exponentially distributed (or more broadly), as we have shown in section 2, then these deviations are power law distributed in the tail with an exponent $\alpha=1/a\nu_c$; (iii) We believe there is no law of large numbers that can adequately describe the entire distribution of the deviations. However, the scale factor (or the typical size of the deviations) can be shown to diverge almost surely if $p\alpha<1$ and to be distributed extremely broadly if $p\alpha>1$. These results demonstrated rigorously have been illustrated 
with numerical simulations. The next section investigates the impact of the propagating uncertainties on the test scores that forecasts receive in earthquake predictability experiments such as CSEP. 

\section{Impact of Magnitude Uncertainties on Model Forecasts and Their Evaluation in Consistency Tests}
\label{LT}

Seismic rates estimated via some probabilistic model at time $t$ in the future as a function of past seismicity are routinely used as direct estimates of a forecast (the expected number of events at time $t$) [e.g. {\it Helmstetter et al.}, 2005; {\it Gerstenberger et al.}, 2005]. In rare cases, forecasts are generated by averaging over Monte Carlo simulations which nevertheless remain parametrized by the estimated seismic rate. Forecasts are therefore either equal to, proportional to or strongly correlated with seismic rate estimates. As a consequence, they suffer from the same fluctuations as seismic rate estimates. 

How important are these spurious fluctuations in the evaluation of forecasts in prediction experiments? Can good forecasts perform poorly solely due to magnitude noise? Can accurate models be rejected by test scores because noisy magnitudes influenced the rate estimation? How sensitive are evaluation procedures to this source of noise? 

We address these questions by designing the following numerical experiment: We test forecasts based on noisy magnitudes against a hypothetical ``reality", chosen as the rate based on exact magnitudes, to see whether the noisy forecasts are rejected. We pretend that the exact seismic rate as calculated from the model equation (\ref{lambdaspt}) in each space-time bin is ``reality" according to which earthquakes actually occur. Since the exact rate is not an integer, we assume that observations are drawn from a Poisson distribution with mean equal to the seismic rate in each bin. We further call the noisy forecasts ``models" ($j=1\dots M$). We then constructed a miniature CSEP testing center which tests these ``models" against ``reality" according to their consistency with the ``observations" in a manner entirely equivalent to the proposed scenario [{\it Schorlemmer et al.}, 2007; {\it Schorlemmer and Gerstenberger}, 2007]. 

The testing center uses the likelihood (L) and number (N) tests as consistency criteria for evaluating forecasts. These are currently used for the five year time-independent forecasts which have already been submitted [{\it Schorlemmer et al.}, 2007]. The test were used in previous studies of forecasting [{\it Kagan and Jackson}, 1994, 1995] and were further explained by {\it Jackson} [1996]. 

We calibrated the numerical experiment to mimic California as much as possible to create realistic conditions under which actual forecast evaluation will take place. This meant choosing a realistic model, realistic model parameters and realistic space-time bins. 

If we find that ``models" (i.e. forecasts based on noisy data) are rejected more frequently by the consistency tests than according to the chosen confidence limit, we can draw the following conclusions: (i) the ``true" model with ``true" parameters may be rejected in a realistic one year test period because noisy magnitude observations affected its forecasts, (ii) the outcomes of the likelihood and number consistency tests are therefore sensitive to observational uncertainties that affected the generation of the forecasts. 

\begin{figure}[tbp]
\begin{center}
\includegraphics[width=16cm]{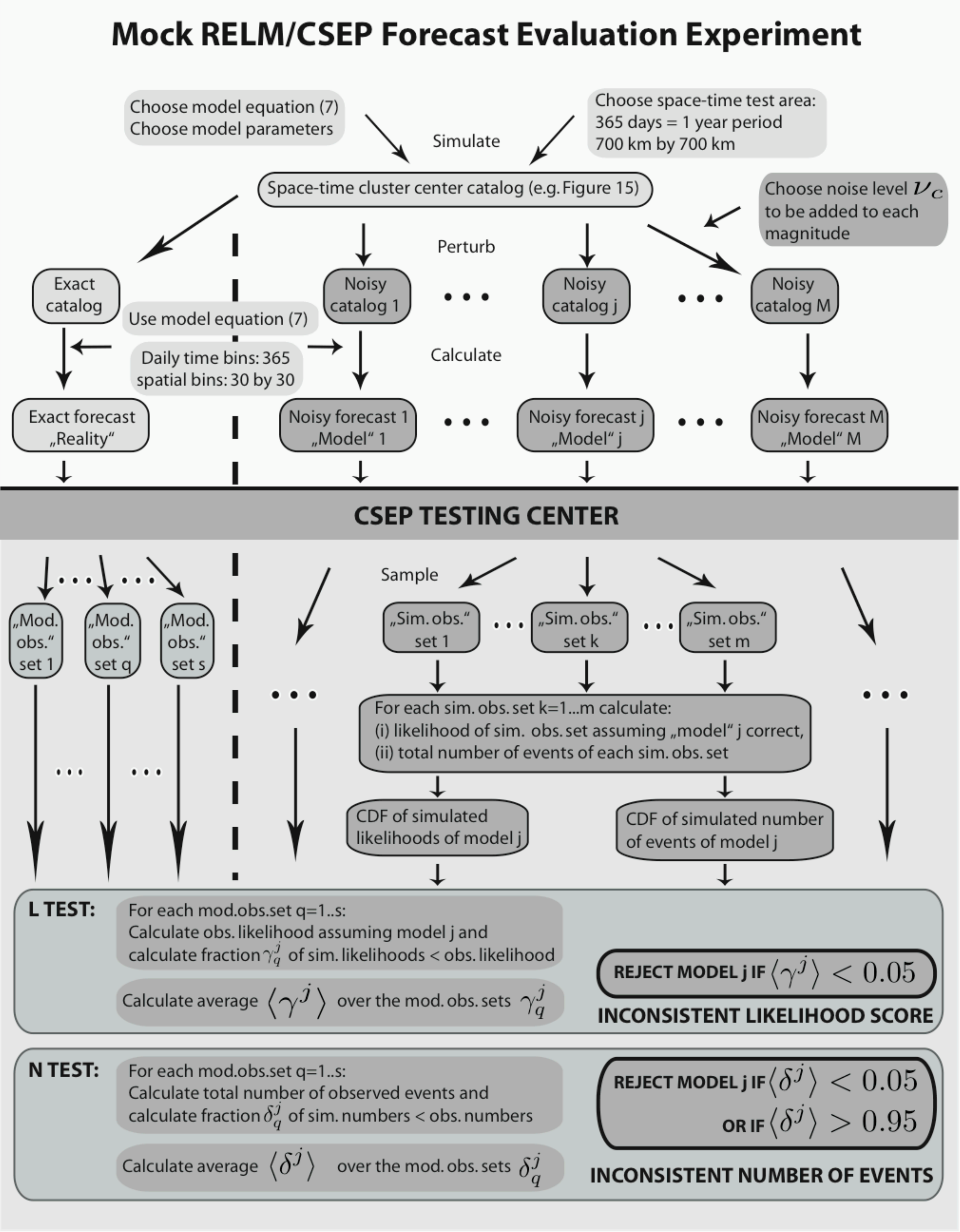}
\end{center}
\caption{\label{chart} Diagram explaining the numerical experiment designed to study the influence of magnitude noise on forecasts and their evaluation in daily earthquake forecast experiments such as RELM or CSEP. We mimic California both in spatial test area and in model parameters and perform the likelihood (L) and number (N) test for daily forecasts over the period of one year using a spatio-temporal Poisson cluster center model that captures essential ingredients of most short-term seismicity models. The abbreviations mod., sim. and obs. stand for modified, simulated and observations, respectively. 
}
\end{figure}

First we describe the simulations and their evaluation by likelihood methods before presenting our results. Our numerical experiment can be separated into two steps: (i) the simulation of the exact seismic rates in each space-time bin and the corresponding noisy forecasts, and (ii) the evaluation of the noisy forecasts with respect to the exact forecast using likelihood tests. Figure \ref{chart} graphically explains our numerical experiment and should be used for reference. 

\subsection{Simulation of Exact Seismic Rates and Noisy Forecasts}

The simulation of the exact seismic rates and the noisy forecasts proceed according to the following steps.

\begin{enumerate}

\item {\bf Choose a model, its parameters and the test area}: The impact of magnitude uncertainties on forecasts will depend on the specific model and potentially its parameters. We chose a spatio-temporal Poisson cluster model defined in equation (\ref{lambdaspt}) to capture the main ingredients of the popular short-term forecasting models. We used parameters and a spatial test area consistent with Californian earthquake data to create realistic conditions (see section \ref{parmchoices} below for parameter values). 

\item {\bf Simulate a ``learning" catalog:} Using a set of parameters and the model (\ref{lambdaspt}), we simulate a ``learning" catalog from which forecasts are to be generated. Because aftershocks of the cluster centers do not trigger their own aftershocks, they do not affect future seismicity and do not need to be simulated for the ``learning" catalog. Only the Poisson cluster center process needs to be simulated. It is a homogeneous space-time Poisson process of constant rate $\lambda_c$ (per unit time and area). Independent magnitudes are drawn from the Gutenberg-Richter law (\ref{GR}). See Figure \ref{simcc1} for an example of a simulated cluster center process. 

\item {\bf Calculate the exact daily seismic rate over one year in each space-time bin $i$:} Divide the test area into spatial cells (bins). Use equation (\ref{lambdaspt}) and the original parameters to calculate the intensity $\lambda(t,{\bf r}|H^c_t)$ for every day over the course of one year at the centers of the spatial cells. Past seismicity (times, locations and magnitudes) enter into equation (\ref{lambdaspt}). The rate in each space-time bin is calculated by multiplying the rate at the center by the area of the spatial cell for simplicity. Pretend these rates are ``reality": the number of actually observed events is drawn from a Poisson distribution with mean equal to this rate (see below). 

\item {\bf Create $M$ noisy replicas of the ``learning" catalog by perturbing the magnitudes:} Generate $j=1 \dots M$ different noisy replicas of the original catalog by adding random noise to each magnitude. The random noise is simulated from the Laplace distribution defined in (\ref{noise}) characterized by the scale parameter $\nu_c$. 

\item {\bf Generate $M$ ``noisy" forecasts based on the $M$ noisy catalogs:} Using equation (\ref{lambdaspt}) and the original parameters, calculate the daily seismic rate for the whole year in the same manner as the exact rate but using the noisy catalogs. Denote these noisy forecasts as ``models" $j=1\dots M$ which we want to compare to the exact seismic rate in a mock CSEP/RELM testing center. 

\end{enumerate}

\subsection{Evaluation of the Noisy Forecasts in a Hypothetical RELM/CSEP Testing Center}

In the evaluation of the ``models" (or noisy forecasts), we follow the recipe of the RELM/CSEP testing procedures, as described below. But first we address an important issue regarding our choice not to use magnitude bins. 

When we simulate the learning catalog above, we must choose a magnitude threshold. We set $m_d=4$ as for the proposed daily forecast competition. Magnitudes for the simulated cluster center events were drawn from the Gutenberg-Richter distribution with $m_d=4$. We then perturbed this catalog according to a symmetric Laplacian distribution. As a result, some of the simulated cluster centers now have magnitudes that are smaller than the magnitude threshold $m_d=4$. We note that their influence is very small in the calculation of the noisy seismic rates (forecasts) with respect to the events above the threshold because of the exponential dependence of the productivity on the difference $\exp(a(m_{i_c}-m_d))$. Rather than re-applying the cut-off $m_d=4$ after the perturbation, we kept the number of events in the learning catalog fixed. We therefore also did not take into account the possibility of magnitudes originally below the threshold becoming ``visible" above the threshold due to the addition of noise. These more realistic conditions should be pursued as a next step. In our scenario, the mean magnitude before and after the perturbation remains constant. Any results in the performance of the forecasts are therefore not due to an overall shift in the magnitude distribution. 

As a consequence, we did not introduce magnitude bins at the testing stage. Rather, we assumed that the seismic rates and the forecasts were for one and the same large magnitude bin. This also meant that we did not need to generate magnitudes of the aftershocks of the cluster centers, neither of the exact catalog nor of the noisy catalogs. This helped the numerical efficiency of the computer program. Furthermore, by not testing the magnitude values, we effectively assumed that all simulated magnitudes were completely consistent with the ``observed" ones, as the sum over the likelihood values of the magnitude bins equals 1. In other words, we are less stringent than the CSEP/RELM center by not rejecting models because of under- or over-predicted magnitudes. 

The CSEP/RELM testing procedure acknowledges the existence of observational uncertainties in the observed earthquake parameters such as magnitude, location or focal parameters. The test center therefore creates so-called ``modified" observations from the actually observed earthquake parameters by sampling parameters that are consistent with the observed ones and their uncertainties. In this way, many alternative ``realities" or observations are created that are consistent with the actual observations and their uncertainties. The forecasts are tested against each alternative reality and the final rejection is based on the average performance of the forecast against all alternative ``realities". 

In our hypothetical center, we should therefore create many ``modified" observations consistent with the actual observation in the same manner. However, as just stated above, we did not actually generate magnitudes of the events and do not test a forecast's ability to predict magnitudes. The magnitude dimension is effectively eliminated from the evaluation phase by assuming that all simulated magnitudes are consistent with the ``observed" ones. 

At the same time, we did generate ``modified" observations in each space-time bin according to the following logic. The CSEP/RELM test center generates so-called ``simulated" observations from a model's forecast that are consistent with the forecast by drawing numbers from a Poisson distribution with mean equal to the forecast in each space-time bin. This assumes that earthquakes are independent and occur according to a Poisson process in each space-time bin. This is not true for the Earth. However, we decided to assume that our ``reality" did indeed occur according to a Poisson process in each bin so that we create favorable conditions for the test to work and to be sure that ``models" were not rejected because they violate this assumption of the test. 

We therefore created many alternative realizations of observed earthquakes that are consistent with the exact seismic rate in each bin according to a Poisson distribution with mean equal to the seismic rate. In analogy with the RELM/CSEP terminology, we called these alternative ``realities" modified observations. For the Earth, these favorable conditions do not hold and may also influence whether a model is rejected. 

With reference to Figure \ref{chart}, the various steps of the RELM/CSEP testing procedure are described below. We follow the notation of {\it Schorlemmer et al.} [2007].  

\begin{enumerate}

\item {\bf For each of the $\{j=1 \dots M \}$ ``models", generate $\{k=1 \dots m\}$ ``simulated" observations}: The RELM/CSEP test center assumes that earthquakes are independent in each space-time bin $i$ and generates Poisson distributed random variables $\hat{w}^j_{k,i}$ with mean equal to the model's forecast $\lambda^j_i$ in that bin. This is repeated $m$-times to generate $\{k=1 \dots m\}$ sets of simulated observations for each model. 

\item {\bf For each of the $M$ ``models", calculate the likelihoods of the $m$ simulated observations}: Assume that the ``model" is true and calculate the likelihood of each of the $\{k=1 \dots m\}$ sets of observations generated from its forecast $\lambda^j_i$ and sum over all space-time bins. These are the simulated likelihood values $\{\hat{L^j_k},k=1\dots m\}$ consistent with a particular model $j$ which will be tested against the likelihood of the observations below. The simulated likelihood of the jth model of the kth observation set is given by: 
\be
\hat{L}^j_k=\sum_i -\lambda^j_i + \hat{w}_{k,i} \log \lambda^j_i - \log ({\hat{w}_{k,i}}!)
\label{simL}
\ee

\item {\bf Generate $\{q=1 \dots s\}$ ``modified" observations (``alternative realities") from the exact seismic rate}: In each space-time bin $i$, generate a ``modified" observation $\tilde{w}^q_i$ by drawing a random variable from a Poisson distribution with mean equal to the exact seismic rate in that bin. Repeat $s$-times to generate $s$ sets of ``modified observations" against each of which each ``model" is tested. 

\item {\bf For each of the $M$ ``models", calculate the likelihoods of the $s$ ``modified" observations}: Assume that the jth ``model" is true and calculate the likelihood $\tilde{L}^j_q$ of observing each of the $\{q=1 \dots s\}$ ``modified'' observations $\tilde{w}^q_i$: 
\be
\tilde{L}^j_q=\sum_i -\lambda^j_i + \tilde{w}^q_i \log \lambda^j_i - \log({\tilde{w}^q_i}!)
\label{obsL}
\ee

\item {\bf For each of the $M$ ``models", calculate the fraction $\gamma^j_q$ of simulated likelihoods $\hat{L}^j_k$ less than each of the $s$ observed likelihoods $\tilde{L}^j_q$}: For each model $j$ and each set $q$ of modified observations, compute the fraction $\gamma^j_q$ of simulated likelihoods $\{\hat{L}^j_k, k=1 \dots m\}$ less than the observed likelihood $\tilde{L}^j_q$. Repeat for each of the $q=1\dots s$ observed likelihoods for each of the $M$ ``models". The fraction $\gamma^j_q$ measures whether an observed likelihood value is consistent with the values expected from a particular ``model". 

\item {\bf For each of the $M$ ``models", calculate the mean fraction $\langle \gamma^j \rangle$ from the $s$ values of $\gamma^j_q$}: Because many ``modified" observations are consistent with our chosen ``reality", we average the fraction $\gamma^j$ over all sets of ``modified" observations for each model (noisy forecast). The mean fraction $\langle \gamma^j \rangle$ measures whether the observed likelihood values fall within the center of the simulated likelihoods given a particular model. 

\item {\bf For each of the $M$ ``models", calculate the simulated total number of events $\hat{N}^j_k$ from each of the $\{k=1 \dots m\}$ simulated observations}: Generate the distribution of the simulated total number of events in the test region consistent with each model. 

\item {\bf Calculate the ``modified" total number of events $\tilde{N}_q$ from each of the $\{q=1 \dots s\}$ modified observations}: For each of the possible ``realities" or modified observations, sum the modified observations over all bins to obtain the modified total number of events $\tilde{N}_q$. 

\item {\bf For each of the $M$ ``models", calculate the fraction $\delta^j_q$ of simulated total numbers of events $\hat{N}^j_k$ less than the modified number of events $\tilde{N}_q$ for each of the $\{q=1 \dots s\}$ modified observations}: Compute the fraction $\delta^j_q$ of simulated numbers of events $\{\hat{N}^j_k, k=1\dots m\}$ less than the qth modified total number of events $\tilde{N}_q$ for each modified observation and each model. The fraction $\delta^j_q$ measures whether the simulated numbers of events are consistent with the observed number. 

\item {\bf For each of the $M$ ``models", calculate the mean fraction $\langle \delta^j \rangle$ from the $s$ values of $\delta^j_q$}: Again, because many modified observations are consistent with the actual observation, average the fraction $\delta^j_q$ over all of its $s$ values for each model (noisy forecast). The mean fraction $\langle \delta^j \rangle$ measures whether the modified observations are on average consistent with the simulations. 

\item {\bf Perform L test: } Reject model $j$ if $\langle \gamma^j \rangle < 0.05$. This indicates that the observed likelihood values are inconsistent with the model. According to Schorlemmer et al. (2007), the L test is one-sided. 

\item {\bf Perform N test: } Reject model $j$ if $\langle \delta^j \rangle < 0.05$ or if $\langle \delta^j \rangle > 0.95$. This indicates that the observed number of events are inconsistent with the model.  

\end{enumerate}

\vspace{.5cm}

\subsection{Test Area and Model Parameter Choices}
\label{parmchoices}

To simulate catalogs and generate exact forecasts, we needed to choose the spatial test area, the number of spatial bins, the parameters of the model and the temporal bins (for which forecasts are issued and evaluated). To mimic a daily forecast competition in California in an earthquake prediction experiment such as RELM, we chose temporal bins of one day and issued forecasts over the course of an entire year always using all information available up to the day of the forecast. We assumed a square spatial area of $700$ km by $700$ km to approximate the size of California. 

To generate synthetic catalogs and forecasts, we needed to decide on a set of parameters $\theta=\{\beta, m_d, \lambda_c, k, a, c, p,\mu\}$ for the model (\ref{lambdaspt}). We decided on $m_d=4$ to mimic the proposed RELM/CSEP daily forecast competition and on $\beta=2.3$. Inverting the parameters of the Poisson cluster process requires knowledge of the entire branching structure (being able to identify which events are cluster centers and which are their dependent aftershocks). Lacking such knowledge, the maximization of the likelihood must be performed over all possible branching structures. Given the complexity, we decided instead to use parameter values based on an ETAS model inversion for southern California of {\it Helmstetter et al.} [2006, model 2 in Table 1] and adjusted them appropriately for our model, magnitude threshold and spatial test area. Their background rate was multiplied by $10^{-2}$ to adjust to the higher magnitude threshold, and multiplied by $3$ for the increased spatial area, resulting in $\lambda_c=0.063$. We also multiplied $k$ by $10^{-2}$ to obtain $k=0.013$. We used their other parameters without change $\{a=0.43 \times \log(10), c=0.0035, p=1.19, \mu=2\}$. We checked that the simulated total number of expected events per year over the entire region given these parameter choices was comparable to yearly rates above M4 in California over the last twenty years (from about 50 up to 250). 

Because of computational limitations, we had to restrict the number of spatial bins to 30 by 30 and the number of perturbations of the original catalog to M=10. However, we were still able to generate $m=1000$ simulated observations for each noisy forecast and $s=1000$ modified observations for the exact forecast. 

In this article, we aim to establish only whether examples exist in which good models are rejected based solely on realistic magnitude uncertainties. A more in-depth study which explores the model, parameter and bin space is certainly required to establish robust confidence limits for testing the consistency of model forecasts with observations. 

\subsection{Simulations and Results}

\begin{figure}[!ht]
\begin{center}
\includegraphics[width=16cm]{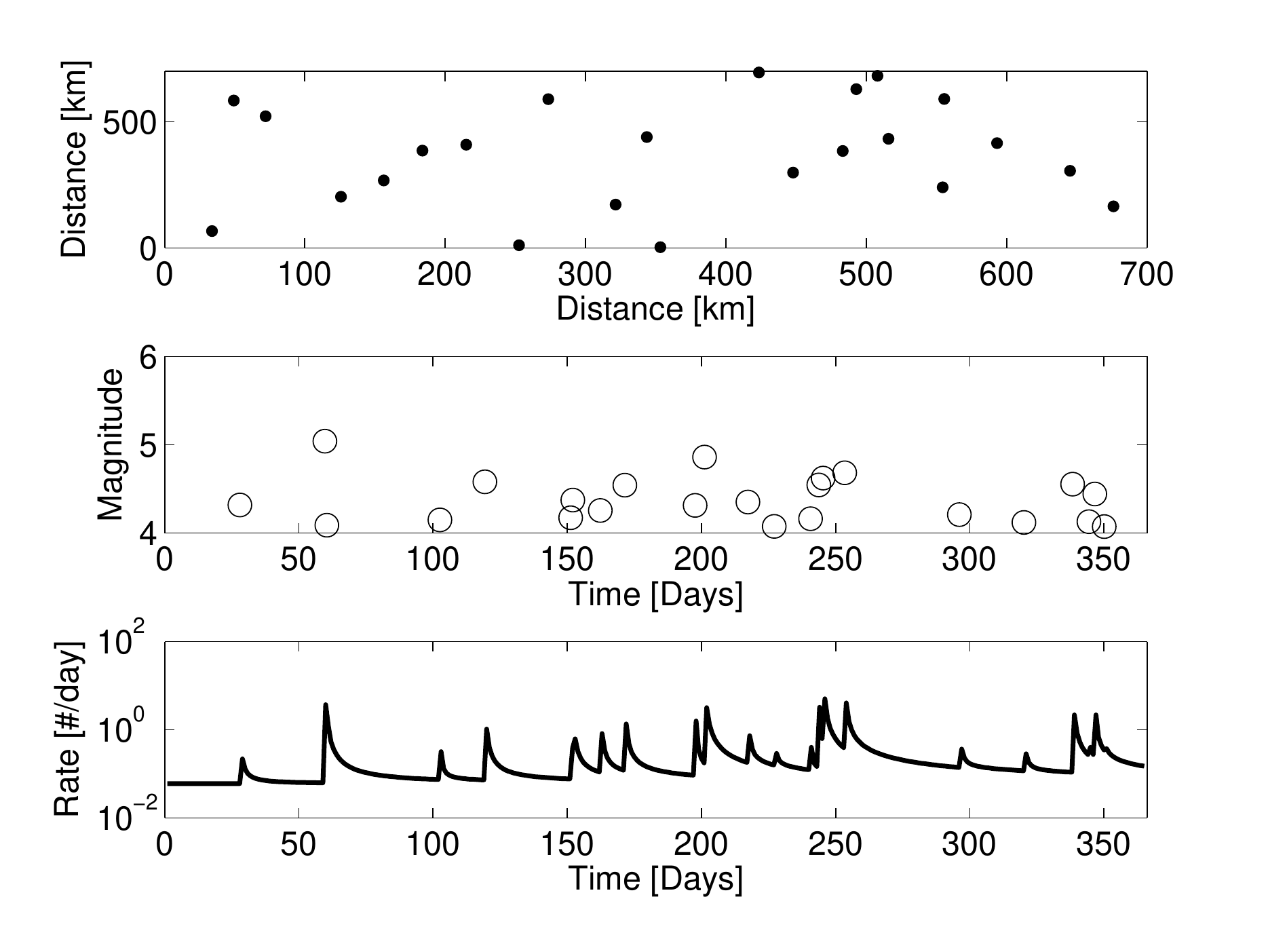}
\end{center}
\caption{\label{simcc1} Example of a simulated cluster center catalog and resulting conditional intensity including aftershocks. Top: spatial distribution. Middle: Magnitudes of cluster centers against time over the course of one year. Bottom: Conditional intensity rate per day calculated using the model equation (\ref{lambdaspt}) but summed over all bins. The rate is used as the ``exact forecast" from which modified observations are generated.
}
\end{figure}

Figure \ref{simcc1} shows an example of a simulated cluster center catalog: the top panel shows the spatial distribution of the cluster centers in the spatial test area $700$ km by $700$ km. The middle panel shows the magnitudes of the cluster centers against time in days over the course of one year. The bottom panel shows the rate for both cluster centers and their aftershocks calculated from the model (\ref{lambdaspt}), summed over all spatial bins. 

We checked that setting the noise level to zero $\nu_c=0.0$ does not result in any ``models" being rejected. Table \ref{tablenu0} shows the results of the mock RELM/CSEP evaluation of the daily forecasts of 10 unperturbed forecasts over the period of one year. None of the ``models" are rejected by the L and N tests. Apart from the mean fractions $\langle \delta^j \rangle $ and $\langle \gamma^j \rangle$, we also calculated their standard deviations $\sigma_\gamma$ and $\sigma_\delta$. Table \ref{tablenu0} shows both fractions to be right in the middle of the simulated distributions, indicating strong consistency between the models and the observations, as should be expected. 

Introducing just a little bit of noise changes the situation. For $\nu_c=0.1$, we found that the N test rejects 2 models at the 90 $\%$ confidence limit (see Table \ref{tablenu1}). We found that the difference between the simulated total number of expected events of a model and the actual expected number based on the exact rate was a good indicator for the model's performance in the N test, as should be expected. In keeping with the statement by {\it Schorlemmer et al.} [2007] that the L test is one-sided, we do not reject models for which $\langle \gamma^j \rangle >0.95$. In this case, the (two-sided) L test would have rejected the same models as the N test. Because rejecting 2 models out of 10 at 90$\%$ confidence may simply be the expected false negative errors, we additionally performed simulations for 50 models. Of these, 8 models were rejected by the N test, none by the L test. Thus in total, 10 models out of 60 models were rejected, indicating slightly higher than expected rejections at 90$\%$ confidence. 

Table \ref{tablenu2} shows the results for simulations with the noise level set to $\nu_c=0.2$. The cluster center process of this simulation is shown in Figure \ref{simcc1} along with the calculated seismic rate according to model (\ref{intensity}). The N test rejects 9 out of 10 models because they over-predict the number of observed events. A two-sided L test would have rejected 3 models, but the one-sided L test does not reject any models. In contrast to the case for $\nu_c=0.1$, the fluctuations strongly impact the forecasts. 

Results for simulations with noise level $\nu_c=0.3$ are summarized in Table  \ref{tablenu3}. Note that the values of $\langle \gamma^j \rangle $ are now fluctuating very strongly. The one-sided L test now rejects one model while the N test rejects 7 models. Clearly, the noisy forecasts are no longer consistent with the observed likelihood values and numbers of events. Note that model 5 predicts more than 10 times the actually expected number of events, reflecting the extreme fluctuations we proved in the previous sections. It is also interesting that the L test rejects model 2, which passes the N test. This shows that the daily expected numbers can fluctuate but in a sense ``average out" over the course of one year, but the likelihood scores of each day keep a ``memory" of the bad predictions. This 
exemplifies the complementary properties of the two tests.

The case $\nu_c=0.5$ is shown in Table \ref{tablenu5}. All models are rejected by the N test, indicating systematically larger forecasts. 

These simulations show that, 
as  $\nu_c$ increases, the models tend to forecast  numbers of events that are larger than the true value: 
the larger $\nu_c$, the larger the effect, and therefore the more probable the rejection by the N test.
This results from the fact that, while the distribution (\ref{noise}) of magnitude errors is found (and assumed in our simulations) to be approximately symmetric, the impact of a magnitude 
error is strongly asymmetric when comparing negative and positive deviations
from the true magnitude, due to the exponential dependence of the productivity law (\ref{rho}).

Because the tests rejected more models for $\nu_c=0.2$ than for $\nu_c=0.3$, we believe that we are not sampling the actual fluctuations with 10 models. More models (perturbed catalogs) are needed to characterize precisely how the confidence limits of the tests are affected. In this article, we showed simply that the stated confidence limits as we would like to interpret them (that the model is inconsistent with 90 $\%$ confidence) are clearly not adequate. 

\subsection{Discussion of Mock RELM/CSEP Predictability Experiment}

The results from the mock RELM/CSEP evaluation of forecasts generated from noisy data against the exact seismic rate based on exact data indicate that noisy forecasts fluctuate wildly and may therefore be rejected by the RELM/CSEP consistency tests. Even conservative levels of magnitude noise of $\nu_c=0.2$ impact forecasts so strongly that they are easily rejected by the likelihood and number tests. In other words, the L and N tests are sensitive to observational uncertainties that entered in the creation of the forecasts. As a consequence, when considering actual results from L and N tests based on comparing a real model with real data, one should keep in mind the possibility that the forecast contains noise which may influence severely the performance of a short-term model. The supposed confidence limits may be misleading as they do not take into account uncertainties in the forecast. Some models may be rejected purely because of forecasts generated from noisy earthquake catalogs, while others may appear to be consistent with the data (more often than expected given the RELM/CSEP confidence limits). 

We do not expect these ``wrongful" rejections to stop if the tests are performed over a longer period of time. Each day is separately scored according to forecast and observations and the daily forecast will always be strongly fluctuating. Extending the evaluation period to two years, for instance, would not solve the problem. 

We emphasize that we have completely isolated the effect of magnitude uncertainties and assumed everything else to be known. We have shown that, in this scenario, magnitude uncertainties lead to strongly fluctuating forecasts. While a comprehensive study of uncertainty in data, parameters, forecasts, observations  and their trade-offs should be encouraged, we expect that there will be no simple formula to ``correct" the forecasts, tests or interpretations. 

Rather, the propagation of data and parameter uncertainties needs to be carefully examined in each specific model and accounted for in the forecasts. The resulting distribution of forecasts can most likely not be captured by one value such as the expectation. In fact, we have shown above that, depending on the noise, we should expect extremely large variations that cannot be represented by one number per bin. 

The L and N tests both assume that earthquakes are independent in each space-time bin and that observations consistent with the forecast are Poisson distributed random variables. In this article, we did not study the implications of the first assumption. It would seem, however, that the assumption of independence will be strongly violated during active aftershock sequences - the days when clustering models can actually be tested on observed clustering. 

But from the standpoint of our results, the second assumption (that observations consistent with a model are Poisson distributed with mean equal to the forecast) may need to be relaxed because of uncertainties in the forecasts, whether due to noisy magnitudes used to generate forecasts, parameter uncertainties or other sources. Instead, models will need to provide the entire likelihood distribution in each space-time bin. Apart from likelihood and number consistency tests, methods for alarm-based earthquake predictions are also equipped to deal with full forecast distributions [see, e.g., {\it Zechar and Jordan}, 2007].

There is a second reason for allowing forecasts to be specified as full distributions. The idiosyncrasies of a model may cause consistent observations to be distributed completely differently than a Poisson distribution. While a certain actual observation may not be consistent with a Poisson distribution given the mean rate of a model forecast, the observation may still be consistent with the model. Specifying the entire distribution of a forecast is computationally much more demanding, but it is the only way to guarantee that forecasts accurately reflect the scientific hypotheses of the model along with all sources of uncertainties involved in the generation of the forecast. 

The power of the ``non-Poisson" L and N tests may appear weaker because less models are rejected, but this is simply a reflection of the potential stochasticity of the model and its real uncertainties. If the tests are adjusted to the distribution of each forecast, then the confidence limits can be interpreted appropriately. 

A trivial way for a model to pass the ``non-Poisson" consistency tests would be to specify extremely broad distributions so that whatever is observed falls into the center of the distribution. Most clustering models are in fact already very broadly distributed so that this may reflect some scientific truth about seismicity. But in case these distributions are too broad, there exists also the likelihood ratio test, which compares models against each other and would be able to reject overly ``dilute" forecasts against peaked ones that are accurate. 

Before concluding, we briefly mention two techniques that seem suitable for generating the entire distribution of forecasts from a model. The first is a simple simulation-based method and is essentially an extension of the method by {\it Rhoades et al.} [1994] from renewal processes to clustering models. The idea is to acknowledge that data and parameters are uncertain and hence distributed and then to repeatedly sample parameters and data randomly from these distributions to generate many forecasts. 

The simulation-based method is simple but computationally expensive. Furthermore, past forecasts of the model are ``thrown away" whenever new observations become available. A second method, data assimilation, provides an optimal and more efficient solution by making use of all available information, including previous model forecasts [see e.g. {\it Kalnay}, 2003]. The goal of data assimilation is to estimate the state of the physical system (and/or parameters) through a statistical combination of the noisy observation and the distributed model forecast according to Bayes' theorem. The state (e.g. past earthquake data and parameters of the model) are sequentially updated through time by correcting the model forecast (the prior) with the observations (the likelihood). 

%Finally, we note that 
\section{Conclusions}

In this article, we analyzed magnitude uncertainties and their impact on seismic rate estimates in short-term clustering models, on their forecasts and on their evaluation in predictability experiments such as RELM or CSEP. In the first part, we quantified magnitude uncertainties. We estimated moment magnitude uncertainties by comparing the estimates for the same events from the CMT and USGS MT catalogs. We found that a double-sided exponential (Laplace) distribution with a scale parameter $0.1$ fit the distribution of the estimate differences significantly better than a Gaussian, reflecting the higher probability of outliers. If the distributions of independent, individual magnitude uncertainties decay much more slowly than a Gaussian, they have at least exponential or fatter-than-exponential tails. We also analyzed MAD values, a measure of magnitude uncertainty, reported by the NCSN in its authoritative region of the ANSS. We found that MAD estimates below $0.1$ may be unreliable. Typical values were between $0.1$ and $0.3$ but outliers occur often. 

Because short-term seismicity models indiscriminately use any listed magnitude in earthquake catalogs for seismic rate projections, inter-magnitude uncertainties reflect the true errors better. We compared the CMT moment magnitudes with the PDE body and/or surface wave magnitudes and found scatter with standard deviations of $0.29$ and $0.26$, respectively. We further found that the NCSN local and coda duration magnitude estimates for the same events fit a Laplace distribution with scale parameter $0.2$ better than a Gaussian (with standard deviation $0.3$). 

The relative lack of available quantitative magnitude uncertainty estimates coupled with their importance for hypothesis testing underscore the need for increased (funding for) data quality assessment and control by network operators.

In the second part, we studied the impact of magnitude noise on seismic rate projections in a simple clustering model that captures the main ingredients of popular short-term models. We proved that seismic rate estimates based on noisy catalog data deviate from their exact rate by power law fluctuations in the tail with exponent $\alpha=(a \nu_c)^{-1}$, where $a$ is the exponent of the productivity law of aftershocks and $\nu_c$ is the scale parameter of the Laplace distribution of the magnitude noise. Thus seismic rate projections and forecasts can fluctuate extremely due to magnitude noise. We further proved that the scale factor $C_{\Delta \lambda}$, which characterizes the typical scale of the fluctuations, remains a random variable and does not converge to a unique, fixed constant. Rather, there are subtle trade-offs between the power law tail, a tendency for the sum of random variables to converge to its stable law (Gaussian or Levy) and the strong quenched disorder due to particular catalog realizations and the stochasticity of the model. 

In the last part, we studied how forecasts based on noisy data would fare in RELM/CSEP predictability experiments. We conducted a numerical experiment in which we constructed a hypothetical testing center and performed a one year test of daily forecasts. We assumed that earthquakes happen according to the seismic rate of a simple clustering model calibrated on an exact catalog data set. We then perturbed the catalog but used the same model to generate forecasts from the noisy data. These were submitted to the mock testing center as ``models" that were tested for the consistency with the hypothetical observations. We found that noisy forecasts were rejected much more frequently than would be expected for a given confidence limit. We concluded that the current RELM number and likelihood consistency tests were sensitive to noisy forecasts and could wrongly reject the ``true" model due to magnitude noise. 

To robustly reject models at specified confidence limit, tests cannot assume that observations consistent with a model are Poisson distributed about its mean rate forecast. To properly capture the idiosyncrasies of each model together with all propagating uncertainties, the forecasts need to specify the entire distribution for each space-time-magnitude bin. Based on our results that forecasts are power law distributed, we expect the deviations from a Poisson distribution to be severe. We noted that data assimilation techniques were particularly useful for propagating entire probability distributions while taking into account all uncertainties. 

\begin{appendix} 

\section{Appendix}
\subsection{The Deviation of the Perturbed Rate from the True Rate as a Sum of Weighted Random Variables}
\label{A1}
This section shows how the deviation of the perturbed rate due to noisy magnitudes from the true rate can be written as a sum over weighted random variables. 

The perturbed rate is given by
\be
\lambda^p(t|H^c_t)= \lambda_c + \sum_{i_c | t_{i_c}<t} {k\ e^{a(m^o_{i_c}-m_d)}\over (t-t_{i_c}+c)^p}
\label{lp}
\ee
while the true rate is given by
\be
\lambda^t(t|H^c_t)= \lambda_c + \sum_{i_c | t_{i_c}<t} {k\ e^{a(m^t_{i_c}-m_d)}\over (t-t_{i_c}+c)^p}
\label{lt}
\ee
where $m^o=m^t+\epsilon$ and $p_{\epsilon}(\epsilon)$ is the distribution of the noise given by
\be
p_{\epsilon} (\epsilon) = { 1 \over 2 \nu_c} e^{\left(  - { |\epsilon | \over \nu_c}   \right)}
\label{noisea}
\ee
Hence, given any catalog realization $H^c_t$ (of cluster centers), the deviation of the perturbed rate from the true rate is: 
\ba
\Delta \lambda (t|H^c_t,\theta)  & & = \lambda^p (t | H^c_t, \theta) - \lambda (t | H^c_t, \theta) \nonumber \\
& & =  \sum_{i_c | t_{i_c}<t} {k\ e^{a(m^o_{i_c}-m_d)}\over (t-t_{i_c}+c)^p} - {k\ e^{a(m^t_{i_c}-m_d)}\over (t-t_{i_c}+c)^p}  \nonumber \\
& & = \sum_{i_c | t_{i_c}<t} {k\ e^{a(m^t_{i_c} + \epsilon_i-m_d)}\over (t-t_{i_c}+c)^p} - {k\ e^{a(m^t_{i_c}-m_d)}\over (t-t_{i_c}+c)^p} \nonumber \\
& & = \sum_{i_c | t_{i_c}<t} {k\ e^{a(m^t_{i_c} -m_d)}\over (t-t_{i_c}+c)^p} \cdot \left( e^{a \epsilon_i} -1  \right) \nonumber \\
& & = \sum_{i_c | t_{i_c}<t} w_i \cdot z_i
\label{rvsum}
\ea
where the last equality expresses the deviation as a sum over a product of two terms: a quenched weight $w_i$ (i.e.. which is fixed for a given catalog but unknowable)
\be
w_i= {k\ e^{a(m^t_{i_c} -m_d)}\over (t-t_{i_c}+c)^p}
\label{weightsapp}
\ee
and a random variable $z_i$
\be
z_i=e^{a \epsilon_i} -1
\label{zapp}
\ee
The weight $w_i$ measures the influence of the ith cluster center according to its magnitude $m^t_{i_c}$ through the productivity law $\rho(m_{i_c})=\exp(a(m^t_{i_c}-m_d))$ and its occurrence time according to the Omori-Utsu law $\phi(t-t_{i_c}) = (t-t_i+c)^{-p}$. The weights $w_i$ thus depend sensitively on the specific catalog realization $\{m^t_{i_c}, t^t_{i_c}\}_{1\le i_c \le N_c}$ and the parameters $\theta$. 

The weights $w_i$ are ``quenched'' or ``frozen'' because they are fixed for a realization but result from a random process. In statistical physics of spin glasses (a similar situation because of the frozen random variables), there are two types of disorder that are treated differently: quenched disorder, where the average is taken over the logarithm of the partition function; and annealed disorder, where the average is taken directly over the partition function. The latter case would correspond in our context to looking at the full distribution of weights, rather than assuming they are fixed. However, we are interested in the fluctuations of the perturbed rate given a fixed catalog.

The random variables $z_i = (\exp(a \epsilon_i) -1)$ modulate the weights due to the magnitude noise $\epsilon_i$. Without noise, $\epsilon=0$ and hence $z_i=0$ so that $\Delta \lambda = 0$. Their distribution is the subject of the next section. 

\subsection{The Distribution of the Random Variables $z$}
\label{A2}
Using the distribution of $\epsilon$ from equation (\ref{noise}), we can determine the distribution of the $z_i$:
\ba
p_z(z) & & = p_{\epsilon}(\epsilon) \ \left|{d\epsilon \over dz} \right| = \left\{ \begin{array} {r@{\quad \quad} l }
		{ 1 \over 2  \nu_c} \ e^{-\epsilon/ \nu_c} \ \left|{d\epsilon \over dz}  \right|,  \hspace{-1cm} & \hspace{0.45cm}Ê0  \le \epsilon < \infty
			\\
		{ 1 \over 2  \nu_c} \ e^{+\epsilon/ \nu_c} \ \left|{d\epsilon \over dz}  \right|, \hspace{-1cm}  & -\infty  < \epsilon < 0 
		\end{array}  \right. \nonumber 
		\\
		& & =  \left\{ \begin{array} {r@{\quad \quad} l } 
		{ 1 \over 2  a \nu_c   (z+1)} \ e^{(-\log(z+1)/a \nu_c  )},   \  & \hspace{0.3cm} 0  \le z < \infty
			\\
		{ 1 \over 2 a \nu_c   (z+1)} \ e^{(\log(z+1)/a\nu_c  )},   \  & -1 < z < 0 
		\end{array} \right.  \nonumber
		\\
		& & =  \left\{ \begin{array} {r@{\quad \quad} l } 
		{ \alpha \over 2  } Ê{1 \over (z+1)^{1+\alpha}},  \ \ \ & \hspace{0.3cm} 0 \leq z < \infty
			\\
		{ \alpha \over 2  }  {1 \over (z+1)^{1-\alpha}},   \ \ \ & -1 < z < 0 
		\end{array} \right.  \hspace{0.1cm} \rm{where} \hspace{0.1cm}Ê\alpha = {1 \over a \nu_c}  \nonumber
		\\
		& & =  {\alpha/2 \over (z+1)^{1 \pm \alpha}}   \ \ \ \ 
		\begin{array} {r@{} l } & (+):  \ \  \ 0 \leq z < \infty  \\ & (-): \   -1 < z < 0 
		\end{array}
\label{pdfz}
\ea

Figure \ref{z} shows a double logarithmic plot of the pdf of the random variable z for several choices of the noise scale parameter $\nu_c=(0.1,0.2,0.3,0.4,0.5)$. We assumed $a=\ln(10) = 2.3$ so that the exponent $\alpha=(4.34, 2.17, 1.45, 1.09, 0.87)$, respectively.

\subsection{Proof of Proposition 1}
\label{A3}
In this section, we prove that
the deviation $\Delta \lambda (t|H_t) $ of the perturbed seismic rates from the true seismic rate due to magnitude noise is a random variable with a distribution having a
power law tail with exponent $\alpha$ and scale factor $C_{\Delta \lambda}$.
Equation (\ref{rvsum}) shows that $\Delta \lambda$ can be written as a finite sum of weighted random variables $z$, where the $z$ are distributed according to (\ref{pdfz}). The proof follows in two steps. First, 
we show that $z$ is regularly varying. Second, we invoke the result that the sum of weighted, regularly varying variables is equally regularly varying in the tail with the same exponent. 
We will frequently refer to the rigorous {\it Jessen and Mikosch} [2006] (hereafter JM), but the definitions and proofs can equally be found in other sources. {\it Sornette} [2004] provides a heuristic and intuitive development of the results we use. 

\vspace{0.5cm}

DEFINITION 2.1 of JM: \textit{ One-dimensional regularly varying random variables $X$ with distribution function $P(X>x)$ are defined by }
\ba
 P(X>x) \sim q' x^{-\alpha} L(x) \hspace{-.3cm} & {\rm and} & \hspace{-0.3cm} P(X \le -x) \sim q'' x^{-\alpha} L(x)  \nonumber \\
& q'+q''=1 &
\label{regvar}
\ea
\textit{where $L$ is a slowly varying function, i.e. $L(cx)/L(c) \to 1$ as $x \to \infty$ for every $c>0$. Condition (\ref{regvar}) is also referred to as a \textit{tail balance condition}. The cases $q'=0$ or $q''=0$ are not excluded. Here and in what follows we write $f(x) \sim g(x)$ as $x\to \infty$ if $f(x)/g(x)\to 1$.}

\vspace{0.5cm}

$z$ is power-law distributed with exponent $\alpha$ for $z>0$ so that the slowly varying function $L(z)$ is simply a constant. For $z<0$, the power law is truncated at $-1$, so that $q''=0$. Hence the tail balance condition is easily verified for the positive tail of $z$. Therefore, the tail of $z$ is regularly varying with index $\alpha$. This concludes the first part of the proof. 

To prove that the finite, weighted sum of regularly varying variables $z$ is also a regularly varying function with the same exponent, we invoke Lemma 3.3 of JM: 

\vspace{0.5cm}
LEMMA 3.3 of JM: \textit{Let $(Z_i)$ be an independently and identically distributed sequence of regularly varying random variables satisfying the tail balance condition (\ref{regvar}). Then for any real constants $\psi_i$ and $m\ge1$}, 
\vspace{0.5cm}
\ba
P(\psi_1 Z_1 + &&\cdots + \psi_m Z_m >x)  \nonumber  \\
 \sim && P(|Z_1|>x) \sum_i^m \left[ q'(\psi_i^+)^\alpha + q''(\psi_i^-)^\alpha \right]  
\label{regvarsum}
\ea
\textit{where $\psi_i^+$ and  $\psi_i^-$ are defined by $P(\psi_i Z_i>x)=P(\psi^+_i Z^+_i>x)+P(\psi^-_i Z^-_i>x)$ where $x^\pm=0 \vee (\pm x)$ (where $\vee$ means ``or").}

\vspace{0.5cm}

In our case, the constants $\psi_i$ are given by the weights $w_i$. Plugging in $q''=0$ (i.e. $q'=1$) from above and using our notation, we have shown that
\be
p(\Delta \lambda) \sim p_z(z) \cdot \sum_i^m (w_i)^\alpha
\label{pdl}
\ee
Denoting $C_{\Delta \lambda}=\sum_i^{N_c} (w_i)^\alpha$, we have shown that
\be
p(\Delta \lambda) \sim {C_{\Delta \lambda}\over (\Delta \lambda)^{1+\alpha}} \hspace{1cm} {\rm for} \hspace{0.3cm} \Delta \lambda \to \infty
\label{pdlf}
\ee
which completes the proof of Proposition 1.

\subsection{Proof of Proposition 2}
\label{A4}

We first state well-established results for a slightly different definition of the scale factor (denoted by $C(t)$, where the time $t$ of evaluation is fixed) for which Proposition 2 can be easily proven (section A4.1). In section A4.2, we consider the more difficult case for our definition of the scale factor ($C(N)$ where the number of cluster centers is fixed). 

\subsubsection{Results for Fixed-Time Scale Factor $C(t)$}
\label{ct}
Recall that the scale factor is defined by: 
\be
C_{\Delta \lambda}=\sum_{i=1}^{N}   {k^\alpha e^{a\alpha(m^t_i-m_d)} \over (t-t_i+c)^{p\alpha}} ~.
\label{jghjblw22}
\ee
When we simulate catalogs, perturb them and calculate the differences between the perturbed rates and the true rate, we are interested in the deviations from the true rate given a fixed number of events. We did not constrain the time $t$ to a fixed value because a more useful and practical result would be the deviations for a fixed number of events. We therefore let $t$ adjust according to when the Nth main shock happened. We then evaluated the rates at time $t=t_N+dt$ just after the Nth main shock. Since the Nth occurrence time is random, $t$ is therefore also random (for different catalogs or realizations of the Poisson process). Let us denote our scale factor by $C(N)$ to stress the fact that the number of main shocks is fixed, not the time $t$. 

On the other hand, a wealth of results is available for the scale factor $C(t)$, for which the time $t$ is fixed and $N$ fluctuates. The scale factor is then defined by:
\be
C(t)=\sum_{i=1}^{N(t)}   {k^\alpha e^{a\alpha(m^t_i-m_d)} \over (t-t_i+c)^{p\alpha}} ~.
\label{jghjblw22b}
\ee
where $N(t)$ is a now random variable for fixed $t$. Now we make the crucial identification of the scale factor $C(t)$ as the intensity of power law shot noise [e.g. {\it Lowen and Teich}, 1990], defined in general by
\be
I(t) = \sum_{j| t_j<t} K_j h(t-t_j)
\label{wewe}
\ee
where $I(t)$ is the ``current" or noise, $t_j$ are Poisson occurrence times with rate $\lambda < \infty$, the $K_j$ are i.i.d. stochastic amplitudes and the ``impulse" function $h(t)= t^{-\delta}$ is an inverse power law function on the interval $[A,B]$ and zero otherwise. These correspond exactly to the fixed-time scale factor $C(t)$, the main shock occurrence times at rate $\lambda_c$, the productivities $k^{\alpha} \exp(-a \alpha(m_j-m_d))$ and the Omori-like decay $(t-t_j+c)^{-p \alpha}$, respectively. Note that in our case $A=c$, $B=\infty$ and $\delta=\alpha p$. 

The cumulants $c_n$ of $I(t)$, which determine the moments of the shot noise, are given by the following equations [Rice, 1945]. For $\delta \neq 1/n$: 
\be
c_n = \lambda \langle K^n \rangle \times {A^{1-n\delta}-B^{1-n\delta} \over n\delta -1}
\label{cn1}
\ee
while for $\delta = 1/n$:
\be
c_n = \lambda \langle K^n \rangle \times \ln(B/A)
\label{cn2}
\ee
The nth cumulant is hence infinite if the nth moment $\langle K^n \rangle$ of the stochastic amplitudes is infinite, if $A=0$ and $\delta \ge 1/n$, or if $B=\infty$ and $\delta \leq 1/n$. 

Since earthquakes cannot have infinite moment (magnitude), their distribution is truncated and hence all moments of the stochastic amplitudes are finite: $\langle K^n \rangle < \infty \ \ \forall n$. However, the productivities are power law distributed up to the truncation with an exponent $\beta/a$ which lies in the range $1<\beta/a<2$. Therefore, the amplitudes fluctuate as if power-law distributed until the sampling actually ``feels" the corner magnitudes. For specific regions of the world, it may take millenia for these corner magnitudes to occur. Therefore, while mathematically all moments of the stochastic amplitude are finite, fluctuations will be power-law like with infinite variance until the upper truncation is actually felt. 

Using results from {\it Rice} [1944] and {\it Lowen and Teich} [1990], we now prove the three elements of Proposition 2 for the fixed-time scale factor $C(t)$: 
\begin{enumerate}
\item First, we show that $C(t)<\infty$ almost surely (a.s.) even as $t \to \infty$ if $p\alpha > 1$. For this, we only need to demonstrate that all cumulants $c_n$ of $C(t)$ given by (\ref{cn1}) or (\ref{cn2}) are finite. We already stated that the moments of the stochastic amplitudes are mathematically finite because the Gutenberg-Richter distribution is truncated. Furthermore, by definition $\delta=p\alpha > 1\geq 1/n \ \forall \ n$, and $A=c>0$. Therefore, $c_n<\infty \ \forall \ n$, which in turn implies that $C(t) < \infty$ a.s. $\forall \ t$. 

\item Second, we show that $C(t)$ diverges a.s. as $t \to \infty$ if $p\alpha < 1$. To prove this, we will bound $C(t)$ from below and show that this lower bound diverges a.s. As noted above, only magnitudes down to $m_d$ are included in the process. Therefore, the smallest productivity is given by $k^{\alpha}$. We create a lower bound for $C(t)$ by replacing all productivities by their lower bound $k^{\alpha}$, i.e. 
\be
C(t) \geq D(t) = \sum_{i=1}^{N_c(t)}   {k^{\alpha} \over (t-t_i+c)^{p\alpha}}
\label{dt}
\ee
Now the process $D(t)$ has moment generating function $Q(s)$ that is equal to equation (A1) in {\it Lowen and Teich} [1990, Appendix A]. There, the authors show that, for $\delta \leq 1$, $Q(s)=1$ for $s=0$ and zero otherwise, so that
$$ \rm{Pr}\{ D_t < x\}=0,    \  \rm{for \ all} \ x< \infty $$
which proves the a.s. divergence of $C(t)$ as $t\to \infty$.

\item Third, in the regime  $p\alpha > 1$ for which $C(t)< \infty $ almost surely, we show that $C(t)$ remains a random variable with a non-degenerate distribution. For this, we only need to state that the variance of $C(t)$ is non-zero as $t \to \infty$. We can actually calculate the variance explicitly, being equal to the second cumulant $c_2$ given by equation (\ref{cn1}): 
\be
\rm{Var}(C_t)={ \lambda_c  \langle K^2 \rangle \over (2 \alpha p -1) c^{2 \alpha p-1}} 
\label{var}
\ee
As stated above, the second moment of the amplitude is mathematically finite. However, if the earthquake catalog under study does not actually sample the upper magnitude cut-off, the variance will behave as if infinite. Thus, not only is $C(t)$ a random variable, it fluctuates wildly. Sampling the corner magnitude may take hundreds to thousands of years even in relatively active regions like California. To get a sense of the numbers, set $\lambda_c = 0.01$,  $\langle K^2 \rangle=1$ and $\alpha p = 2$ for simplicity, neglecting for a moment the large second moment of $K$. For these values, the variance is on the order of $10^6$, far larger than typical earthquake rates of e.g. 1 per day above $m_d=4$ in California. This completes the proof of Proposition 2 for the fixed-time scale factor $C(t)$.
\end{enumerate}

In fact, more results are known about the statistical properties of $C(t)$ [e.g. {\it Lowen and Teich}, 1990, Figure 3, and references therein]. If $A>0$ and $\delta>1$ so that cumulants exist (assuming the stochastic amplitudes have finite moments), then in the limit of infinite Poisson driving rate $\lambda \to \infty$, the intensity $C(t)$ is distributed according to a Gaussian with mean equal to the first cumulant and variance equal to the second cumulant. Even in this limit (which is {\it not} directly relevant to earthquakes since there $\lambda$ is small) the variance remains huge. Furthermore, if $A=0$ and $\delta>1$, the distribution of $C(t)$ is Levy-stable for all Poisson rates with exponent $1/\delta$. Since for earthquakes, $A=c$ is very small, we expect the distribution of $C(t)$ to be close to Levy with exponent $1/\alpha p= a \nu_c / p $. For reasonable values $a=2.3$, $\nu_c=0.2$ and $p=1.2$, this results in an extremely small Levy exponent 0.4.

\subsubsection{Results for Fixed-Number Scale Factor $C(N)$}
\label{cn}

Results do not seem widely established for the fixed-number scale factor. Note, however, that both the mean number and the variance of the number of events in a Poisson process diverge as $t \to \infty$. The higher moments of degree $n$ of the Poisson process are Touchard polynomials of degree $n$ of the variable $\lambda t$. For $t \to \infty$, they also diverge. Furthermore, the probability of having a finite number k of events in an infinite interval Pr$\{N(0,T]=k\}= (\lambda T)^k \exp(-\lambda T) / k! $ is zero for $T \to \infty$. And vice versa, the probability of having an infinite number of events $k=\infty$ in a finite interval $T$ is equally zero. Therefore, the limit $t \to \infty$ and $N \to \infty$ are equivalent so that we expect the same results to hold for $C(N)$ as for $C(t)$. Without a more formal statement of this equivalence, however, we proceed to consider separately $C(N)$ as $N \to \infty$. 

Let us rewrite expression (\ref{jghjblw}) for $C(N)$ explicitly as
\be
C(N)=\sum_{i=1}^{N}   {k^\alpha e^{a\alpha(m^t_i-m_d)} \over (t_N-t_i+c')^{p\alpha}} ~.
\label{jghjblw22c}
\ee
where $c'=c+dt$ is a constant since $t=t_N+dt$. We bound $C(N)$ from below and from above by noting that
\be
k^\alpha  \leq k^\alpha e^{a\alpha(m^t_i-m_d)} \leq k^\alpha e^{a\alpha (M-m_d)}~.
\label{www}
\ee
since $m_d \leq m^t_i \leq M$, 
where $M$ is an upper magnitude bound, which always exists due to the finiteness of the Earth. Thus, 
\be
k^\alpha \sum_{i=1}^{N_c}   { 1\over (t_N-t_i+c')^{p\alpha}}  \leq 
C(N) \leq k^\alpha e^{a\alpha (M-m_d)}
\sum_{i=1}^{N_c}   {1 \over (t_N-t_i+c')^{p\alpha}} ~.
\label{jghjblw2ww2}
\ee

\begin{enumerate}
\item First, we show that, when $p\alpha > 1$, the scale factor $C_{\Delta \lambda}$ converges to a finite value as $N_c \to \infty$ under the assumption of an arbitrary small but finite minimum time interval $0<\tau_{min}<<1/\lambda$ between events. In this case, we can further bound the right-hand-side of equation (\ref{www}) from above by replacing the intervals $t_N-t_i$ by $(N-i) \tau_{min}$:
\be
C(N) \leq k^\alpha e^{a\alpha (M-m_d)} \sum_{i=1}^{N_c}   {1 \over ((N-i) \cdot \tau_{min}+c')^{p\alpha}}
\label{weo}
\ee
The sum in (\ref{weo}) is in turn bounded from above by the Riemann zeta function $\zeta(\alpha p)=\sum_j^\infty 1/j^{\alpha p}$, which converges absolutely for $\alpha p>1$. This completes the proof that the scale factor $C_{\Delta \lambda}$ converges to a finite value as $N_c \to \infty$ if $p\alpha > 1$.

\item Second, we show, when $p\alpha \leq 1$, that (i) the expectation of the scale factor $C(N)$ diverges as $N_c \to \infty$ using Jensen's inequality and (ii) $C(N)$ diverges under the assumption of an arbitrarily large but finite maximum interval $\tau_{max}$ between events. In the latter case (ii), we can bound $C(N)$ from below by replacing the intervals $t_N-t_i$ by $(N-i) \tau_{max}$:
\be
C(N) \geq k^\alpha  \sum_{i=1}^{N_c}   {1 \over ((N-i) \cdot \tau_{max}+c')^{p\alpha}}
\label{weo_b}
\ee
for which the right-hand-side diverges for $p \alpha <1$ so that $C(N)$ diverges. In the other part (i), we show that the expectation of $C_{\Delta \lambda}$ diverges using Jensen's inequality [e.g. {\it  Durrett}, 2005]. Jensen's inequality theorem states that for any convex function $g(x)$ (i.e., with non-negative second derivative $g''(x) \ge 0$, $ \forall  \ x$, if the second derivative exists) and for any random variable $\xi$ with finite expectation, the following inequality holds true:
\be
{\rm E}[g(\xi)] \ge g ({\rm E}[\xi])~,
\label{jensen}
\ee
where ${\rm E}[x]$ denotes the expectation of the random variable $x$. 
The equality sign holds true in (\ref{jensen}) only for a degenerate distribution of $\xi$. Now we use $g(x) = 1/x^{\alpha p}$, having checked that its second derivative is positive $g''(x)>0$, and let the random variable be $\xi=t-t_i+c$. Using Jensen's inequality:
\be
{\rm E}\left[{1\over (t-t_i+c)^{\alpha p}}\right] \ge {1\over ({\rm E}[t-t_i+c])^{\alpha p}}~.
\label{jensen2}
\ee
Now if the $t_i$'s are assumed to be a stationary sequence (not even necessarily Poissonian), then $E[t-t_i+c]= const\cdot (N-i) +c$. Using this argument on each term in the sum leads to
\be
{\rm E}\left[ \sum_{j=1}^N {1\over (t-t_i+c)^{\alpha p}}     \right] \ge \sum_{j=1}^N {1 \over (const\cdot (N-j) +c)^{\alpha p}}
\ee
But the term on the right hand side diverges $\to \infty$ as $N \to \infty$ for $\alpha p < 1$ so that the expectation of the scale factor also diverges. This proves the second element of Proposition 2.

\item Finally, in the regime  $p\alpha > 1$
for which $C(N)$ converges almost surely
to a finite value as $N \to \infty$, we show that $C(N)$ remains a random variable
dependent on the specific catalog distributed according to a non-degenerate distribution. We rewrite (\ref{jghjblw22c}) as
\be
C(N)=\sum_{i=1}^{N}  \omega_i  X_i~,
\label{hn;krhg}
\ee
where $\omega_i \equiv  1 /(t-t_i+c)^{p\alpha}$ and $X_i \equiv k^\alpha e^{a\alpha(m^t_i-m_d)}$.
The scale factor $C(N)$ is rewritten in (\ref{hn;krhg}) as a randomly weighted sum of i.i.d. random variables $X_i$, where the random weights are functions of the random occurrence times and the random variables are the magnitude-dependent productivities. The weights are non-identically distributed and dependent while the $X_i$ are i.i.d. We will show that for any fixed configuration of occurrence times and random $X_i$, $C(N)$ remains distributed. For $p\alpha > 1$, we have shown in 1. that $W_N \equiv \sum_{i=1}^{N}  \omega_i < \infty$ for $N \to \infty$. We then use the
 result quoted from {\it Jamison et al.} [1965]: ``[if the sum $W_N$ of the weights converges, then] $C(N)/W_N$ [the normalized weighted sum] either fails to converge in probability or converges almost surely to a non-degenerate limit." In plain words, in the latter case, this means that the random variable $S_n/W_n$ remains distributed according to a non-degenerate probability distribution, even in the limit $N \to \infty$. Thus, in both cases, the variance of the scale factor remains non-zero. The intuition behind this result is that the convergence of the weights ensures that there are only a finite number of terms in the infinite sum that contribute to it. This implies that, notwithstanding the existence of an infinite number of contributions, the sum remains a random variable controlled by a finite number of them. This completes the proof that the scale factor does not converge to a unique constant (a degenerate limit) when the exponent $p/(a \cdot \nu_c)>1$. 
\end{enumerate}

\end{appendix}

\begin{acknowledgments}
We acknowledge the Advanced National Seismic System, the Northern California Seismic Network, the US Geological Survey, the NEIC/PDE and the (Harvard) CMT project for the earthquake catalogs used in this study. We thank Margaret Hellweg of the Berkeley Seismological Laboratory and David Oppenheimer of the USGS for information about magnitude uncertainties. We gratefully acknowledge insightful inputs from Kayo Ide, David Jackson, Vladilen Pisarenko, Alexander Saichev and Jeremy Zechar. This work has been partially supported by the grant number NSF ATM-0327900. M.J.W. gratefully acknowledges financial support from a NASA Earth System Science Graduate Student Fellowship. 
\end{acknowledgments}

\end{article}

\newpage

\begin{table}
\begin{center}
\begin{tabular}{|c|c|c|c|c|c|}
\hline
``Model" & E(N) & $\langle \gamma \rangle$ & $\sigma_{\gamma}$ & $\langle \delta \rangle$ & $\sigma_{\delta}$ \\
\hline
%0.0 
 %143.61 
\hline
1& 143.61 &0.4996& 0.2835 &0.5361& 0.2849 \\
\hline
 2&143.61& 0.5106 &0.2795 &0.5162 &0.2866 \\
\hline
 3&143.61 &0.4895& 0.2887 &0.5279 &0.2866 \\
\hline
4& 143.61& 0.5051& 0.2791& 0.5274 &0.2841 \\
\hline
5& 143.61 &0.5028& 0.2837 &0.5185 &0.2910 \\
\hline
6& 143.61& 0.5138& 0.2864& 0.5177& 0.2870 \\
\hline
 7&143.61 &0.4913 &0.2897 &0.5350& 0.2975 \\
\hline
8& 143.61& 0.5131& 0.2882& 0.5157 &0.2921 \\
\hline
9& 143.61 &0.4948 &0.2866 &0.5136 &0.2870 \\
\hline
 10&143.61& 0.5110& 0.2848& 0.5154& 0.2846 \\
\hline
\end{tabular}
\end{center}
\caption{Results of the mock RELM/CSEP experiment of daily forecasts over the period of one year for $\nu_c=0.0$: We checked that ``models" are not rejected by the tests when the data is exact and no noise is present. ``Models" are forecasts generated from equation (\ref{lambdaspt}) using a noisy cluster center process which was perturbed from the original one by adding random noise of scale $\nu_c$ to the magnitudes. The first column contains different perturbations of the original catalog corresponding to different forecasts or ``models" (which, for $\nu_c=0.0$, are all equal). The second column is the total expected number of events obtained by summing all daily forecasts over all spatial bins over the one year period. The expected number of events of the original catalog was $143.61$. The third column shows the fraction $\langle \gamma \rangle$ of the $m$ simulated likelihoods less than the observed likelihood, averaged over all $s$ modified observations. The fourth column shows the standard deviation of the $\gamma$ values. The fifth column shows the fraction $\langle \delta \rangle$ of the $m$ simulated numbers of events less than the observed number of events, averaged over all $s$ modified observations. The sixth column shows the standard deviation of the $\delta$ values. 
}
\label{tablenu0}
\end{table}

\begin{table}
\begin{center}
\begin{tabular}{|c|c|c|c|c|c|}
\hline
``Model" & E(N) & $\langle \gamma \rangle$ & $\sigma_{\gamma}$ & $\langle \delta \rangle$ & $\sigma_{\delta}$ \\
\hline
 %0.1 
 %136.31 
\hline 
1& 117.31& 0.3361& 0.2647 &0.9011& 0.1552 \\
\hline
2& 187.36& 0.9546&0.0880 &0.0018 &0.0109 \\
\hline
3& 150.23& 0.5262 &0.2921 &0.2038 &0.2106 \\
\hline
4& 145.38 &0.5981 &0.2729& 0.2858 &0.2461 \\
\hline
5& 665.09&1.0000& 0.0000& 0.0000 &0.0000 \\
\hline
6& 143.25& 0.4935&0.2870 &0.3321& 0.2574 \\
\hline
7& 139.40 &0.4826 &0.2812 &0.4335& 0.2749 \\
\hline
8& 162.28& 0.6158 &0.2705 &0.0627 &0.1054 \\
\hline
9& 124.38 &0.2920 &0.2573& 0.7817 &0.2321 \\
\hline
10& 130.70&0.3203& 0.2639& 0.6454 &0.2759 \\
\hline
\end{tabular}
\end{center}
\caption{Same as Table \ref{tablenu0} but now perturbing the original catalog from which forecasts (``models") are generated by introducing noise $\nu_c=0.1$. The total expected number of events based on the exact forecast was $E(N)=136.31$. The L test does not reject any models while the N test rejects 2 models. 
}
\label{tablenu1}
\end{table}

\begin{table}
\begin{center}
\begin{tabular}{|c|c|c|c|c|c|}
\hline
``Model" & E(N) & $\langle \gamma \rangle$ & $\sigma_{\gamma}$ & $\langle \delta \rangle$ & $\sigma_{\delta}$ \\
\hline
\hline
%0.2 
  %95.81 
 1&124.11 & 0.8077 &0.2072 &0.0292 &0.0577 \\
\hline
2& 120.29& 0.5917& 0.2845& 0.0495 &0.0892 \\
\hline
3& 135.62& 0.8260 &0.1984 &0.0042 &0.0133 \\
\hline
4&  98.97 &0.3816& 0.2730& 0.4288 &0.2813 \\
\hline
5& 156.55& 0.9810& 0.0547 &0.0000& 0.0004 \\
\hline
6& 170.75 &0.9949& 0.0222& 0.0000 &0.0000 \\
\hline
7& 246.27 &1.0000 &0.0000 &0.0000& 0.0000 \\
\hline
8& 128.59 &0.8307 &0.2033 &0.0150& 0.0369 \\
\hline
9& 121.60& 0.7972& 0.2208 &0.0367 &0.0713 \\
\hline
10& 139.99& 0.8710& 0.1719& 0.0020& 0.0074 \\
\hline
\end{tabular}
\end{center}
\caption{Same as Table \ref{tablenu1} but now perturbing with stronger noise $\nu_c=0.2$. The total expected number of events based on the exact forecast was $E(N)=95.81$. The L test does not reject any models while the N test rejects 9 models. 
}
\label{tablenu2}
\end{table}

\begin{table}
\begin{center}
\begin{tabular}{|c|c|c|c|c|c|}
\hline
``Model" & E(N) & $\langle \gamma \rangle$ & $\sigma_{\gamma}$ & $\langle \delta \rangle$ & $\sigma_{\delta}$ \\
\hline
%0.3 
 %186.65 
\hline 
1& 329.28 &1.0000 &0.0007 &0.0000& 0.0000 \\
\hline
2& 179.76 &0.0406& 0.0714 &0.6250 &0.2806 \\
\hline
3& 398.07& 1.0000 &0.0000 &0.0000 &0.0000 \\
\hline
4& 162.82 &0.0932 &0.1311& 0.8936& 0.1533 \\
\hline
5&3258.11 &1.0000 &0.0000 &0.0000 &0.0000 \\
\hline
6& 249.14 &0.8388& 0.1896 &0.0015& 0.0058 \\
\hline
7& 323.39 &0.9863 &0.0418& 0.0000&0.0000 \\
\hline
8& 204.90& 0.1300& 0.1600 &0.1853 &0.2159 \\
\hline
9& 257.31& 0.8528& 0.1830& 0.0007& 0.0021 \\
\hline
10& 288.11& 0.6957& 0.2588& 0.0000 &0.0000 \\
\hline
\end{tabular}
\end{center}
\caption{Same as Table \ref{tablenu1} but now perturbing with stronger noise $\nu_c=0.3$. The total expected number of events based on the exact forecast was $E(N)=186.65$. The L test rejects 1 model while the N test rejects 7 models. 
}
\label{tablenu3}
\end{table}

\begin{table}
\begin{center}
\begin{tabular}{|c|c|c|c|c|c|}
\hline
``Model" & E(N) & $\langle \gamma \rangle$ & $\sigma_{\gamma}$ & $\langle \delta \rangle$ & $\sigma_{\delta}$ \\
\hline
%0.5 
  %92.15 
\hline
 1&349.73& 1.0000 &0.0000 &0.0000& 0.0000 \\
\hline
2&11715.37 &1.0000& 0.0000& 0.0000& 0.0000 \\
\hline
 3&691.43& 1.0000& 0.0000& 0.0000& 0.0000 \\
\hline
4&9150.17& 1.0000& 0.0000& 0.0000& 0.0000 \\
\hline
5&1248.60 &1.0000& 0.0000& 0.0000& 0.0000 \\
\hline
6& 341.29& 1.0000& 0.0000& 0.0000& 0.0000 \\
\hline
7& 457.51& 1.0000& 0.0000& 0.0000& 0.0000 \\
\hline
8& 645.55& 1.0000& 0.0000& 0.0000& 0.0000 \\
\hline
9&5593.47& 1.0000& 0.0000& 0.0000& 0.0000 \\
\hline
10& 217.19& 0.4032& 0.2537& 0.0000& 0.0000 \\
\hline
\hline
\end{tabular}
\end{center}
\caption{Same as Table \ref{tablenu1} but now perturbing with stronger noise $\nu_c=0.5$. The total expected number of events based on the exact forecast was $E(N)=92.15$. The L test does not reject any models while the N test rejects all 10 models. 
}
\label{tablenu5}
\end{table}

%\clearpage

%FIGURE 1

%\clearpage

%FIGURE 2

%FIGURE 3

%\clearpage

%FIGURE 4

%\newpage
%\clearpage

%FIGURE 5

%FIGURE 6

%\clearpage

%FIGURE 7

%\clearpage

%FIGURE 8

%\clearpage

%FIGURE 9

\clearpage

%FIGURE 10

%\clearpage

%FIGURE 11

%\clearpage

%FIGURE 12

%\clearpage

%FIGURE 13

%\clearpage

%FIGURE 14

%\clearpage

%FIGURE 15

\end{document}